\documentclass[10pt]{iopart}

\usepackage[utf8]{inputenc}

\expandafter\let\csname equation*\endcsname\relax
\expandafter\let\csname endequation*\endcsname\relax
\usepackage{amsmath}

\usepackage{iopams}
\usepackage{bm}
\usepackage{graphicx}
\usepackage{color}
\usepackage{physics}
\usepackage{siunitx}
\usepackage{bbold}

\DeclareSIUnit\angstrom{\text {Å}}

\definecolor{Blue}{rgb}{0.3,0.3,0.9}
\definecolor{Red}{rgb}{0.9,0.3,0.3}
\definecolor{Green}{rgb}{0.3,0.6,0.3}
\definecolor{Black}{rgb}{0.0,0.0,0.0}

\newcommand{\ce}[1] {$\mathrm{#1}$}
\renewcommand{\c}[1]{\mathcal{#1}}

\newcommand{\ham}{\mathcal{H}}
\newcommand{\dt}{\mathop{dt}}

\newcommand{\id}{\sigma_0}
\newcommand{\sx}{\sigma_x}
\newcommand{\sy}{\sigma_y}
\newcommand{\sz}{\sigma_z}

\begin{document}
\title{Radiation-induced dynamical formation of Floquet-Bloch bands in Dirac Hamiltonians 
}
%
\author{Yuriko Baba$^{1,2}$, Vanessa Junk$^3$, Wolfgang Hogger$^3$, Francisco Dom\'{\i}nguez-Adame$^2$, Rafael A. Molina$^4$, Klaus Richter$^3$}
\address{$^1$ Condensed Matter Physics Center (IFIMAC), Universidad Autónoma de Madrid, E-28049 Madrid, Spain}
\address{$^2$ GISC, Departamento de F\'{\i}sica de Materiales, Universidad Complutense, E-28040 Madrid, Spain}
\address{$^3$ Institute of Theoretical Physics, University of Regensburg, 93040 Regensburg, Germany}
\address{$^4$ Instituto de Estructura de la Materia IEM-CSIC, Serrano 123, E-28006 Madrid, Spain}

\ead{yuriko.baba@uam.es}

\date{\today}

\begin{abstract}

Recent experiments, combing ultrafast strong-field irradiation of surfaces with time- and angle-resolved photoemission spectroscopy, allow for monitoring the time-dependent charge carrier dynamics and the build-up of transient sidebands due to the radiation pulses.
While these structures are reminiscent of Floquet-Bloch bands, standard Floquet theory is not applicable since it requires a strictly time-periodic driving field.
To study the emergence and formation of such sidebands, i.e.~to provide a link between common Floquet physics and dynamical mechanisms underlying short driving pulses, we consider a generalization of Floquet theory, the so-called $t-t^{\prime}$ formalism.
This approach naturally extents Floquet theory to driving field amplitudes with a superimposed envelope shape.
Motivated by experiments we study 2D Dirac Hamiltonians subject to linearly and circularly polarised  light waves with a Gaussian field envelope of a few cycles. 
For these Floquet-Bloch Hamiltonians we study the evolution of their Floquet-Bloch spectra, accompanied by a systematic analysis of the time-dependent (sideband) transitions. 
We show that sideband occupation requires circularly polarized light for linear Dirac systems such as graphene, while for Dirac models with trigonal warping, describing 
surface states of topological insulators such as \ce{Bi_2 Se_3}, both linearly and circularly polarised pulses induce sideband excitations.

\end{abstract}

\maketitle

\section{Introduction} \label{sec:Intro}

External time-dependent electromagnetic fields have emerged as a fruitful tool for controlling quantum materials~\cite{Aoki2014,Basov2017}. The idea of controlling a quantum lattice system by a strong periodic external field dates back to the proposal of Dunlap and Kenkre in 1986~\cite{Dunlap1986}. In their seminal work, a charged particle in a 1D tight-binding model was studied under the effect of a sinusoidal driving. The external field delocalises the state, although a localised phase could be induced by varying the magnitude and frequency of the external electric field~\cite{Dunlap1986}. The occurrence of a localised phase by tuning the external field, coined  coherent destruction of tunnelling~\cite{Grossmann1991}, represents an early example of quantum control through external driving, for a review see~\cite{Grifoni1998}. This effect was first observed in Bose-Einstein condensates in a shaken optical lattice~\cite{Lignier2007, Sias2008} where the suppression of tunnelling for large frequencies was found to correspond to a Bessel function law.

The idea of controlling the response of the system by the external field introduced in Dunlap's work has evolved to the more general concept of Floquet engineering~\cite{Basov2017, Oka2019, Rudner2020}, where the effective renormalisation of the parameters of a quantum system is tuned by the external driving, especially in the high-frequency regime. Floquet engineering has been successfully implemented in ultracold atoms, where the modulation is introduced by shaking the underlying optical lattice~\cite{Eckardt2017, Goldman2016,Dotti2024}.
The realisation of the Floquet engineering technique in solid state systems is a challenging but fruitful, expanding field~\cite{Wang2013b, Mahmood2016, McIver2019}. In this case, the modulation is performed by external electromagnetic radiation. To achieve the high driving amplitudes required, ultrashort laser pulses are commonly used. This leads to short-lived Floquet phases and difficulties in detecting them.

Recently, the field of Floquet condensed matter has been revisited with the aim of controlling the topological properties of the systems. Indeed, a topological phase can be induced by applying circularly polarised light to graphene, leading to a quantum Hall insulator~\cite{Oka2009, Kitagawa2011, Oka2019, McIver2019}. Similarly, a Floquet topological insulator was proposed in semiconductor quantum wells, where the topological phase can be tuned by a frequency greater than the bandgap~\cite{Lindner2011, GomezLeon2013}. Experimentally, these phases have been achieved in optical lattices by a synthetic topological gauge~\cite{Goldman2016}, while the light-induced quantum anomalous Hall effect was first reported in graphene in \cite{McIver2019}.

On the other hand, time-resolved ARPES measurements have allowed the imaging of Floquet-Bloch states at the surface of the topological insulator  \ce{Bi_2 Se_3}~\cite{Mahmood2016,Wang2013b}. In the first experiments, an intense ultrashort mid-infrared pulse with energy below the bulk band gap dresses the Bloch states by creating gaps due to hybridisation of the surface states bands~\cite{Wang2013b}. Indeed, together with the accompanying replicas of the original bands, avoided crossings in the momentum space appear due to the breaking of symmetries by the driving, as predicted by theory~\cite{Lindner2011, Farrell2016}. Thanks to the striking time resolution of recent ARPES measurements~\cite{Reimann2018, Ito2023, Reimann2023}, the dynamical formation of sidebands at the surface states of topological insulators has been observed even with subcycle precision~\cite{Ito2023}. 

The experimental observation of the bands within a cycle naturally raises the question of how the Floquet structure is actually built up from a theoretical perspective. This paper aims to answer this question by analysing the case of non-periodic driving fields. To this end, the $t-t^{\prime}$ method is employed to study the time-dependent Floquet band structure~\cite{Howland1974,Peskin1993,Grifoni1998,Drese1999, Holthaus2015}. The $t-t^{\prime}$ formalism considers two time scales, namely the envelope timescale and the period related to the frequency, which should be much shorter than the envelope for the method to work efficiently. In this way, the pulse driving is described using an instantaneous Floquet basis, with the pulse amplitude as a decoupled parameter. The formalism has been shown to successfully describe strong pulses in two-level systems~\cite{Ikeda2022}. Here we consider the case of an effective 2D Dirac Hamiltonian describing the surface states of 3D topological insulators, as they are probed in  ultrafast experiments~\cite{Reimann2023,Schmid2021}.

The paper is structured as follows. In \sref{sec:FQNonPeriodic}, the standard Floquet formalism is adapted to non-periodic pulses, starting from a brief review of the Floquet formalism for periodic systems in \sref{sub:FQPeriodic}. In \sref{sub:ttpMod}, the $t-t^{\prime}$ formalism is introduced  and then applied to Bloch Hamiltonians in \sref{sub:ttpBloch}. The comparison of this formalism with the results from the direct solution of the time-dependent Schr\"{o}dinger equation is briefly discussed in \sref{sub:ttpvsUt}. Next, the $t-t^{\prime}$ formalism is applied to 2D Dirac models in \sref{sec:Applications}. First, the linear Dirac cone is considered in \sref{sub:LinDirac} for linear and circularly polarised Gaussian pulses. Then the case of surface states of \ce{Bi_2 Se_3} is discussed in \sref{sub:Bi2Se3}. The model considered can also be employed to describe  topological insulators such as \ce{Bi_2 Te_3} and  \ce{Sb_2 Te_3}.
Finally, \sref{sec:Conclusion} summarises our main results. 

\section{Floquet theory for non-periodic drivings} \label{sec:FQNonPeriodic}

The interest in describing non-periodic driving is twofold. On the one hand, in condensed matter systems the strong driving regime is usually accessible only with short pulses. On the other hand, the recent time-resolved ARPES measurements open the way to experimentally access
how  photo-dressing of electrons is dynamically built up. Indeed, in these experiments the pump pulses are typically of the order of $10\,\si{\tera\Hz}$ (i.e. with $T^\mathrm{pulse} \sim 10 \,\si{\femto \s}$) and thus the subcycle regime can be accessed with the typical $\si{\femto \s}$ resolution of the time-resolved ARPES setups~\cite{Rohde2016, Reimann2018, Reimann2023}. The usual time dependence of the driving resembles a Gaussian-like pulse. Hence, the Floquet basis does not naturally arise from the long time limit. Nevertheless, a Floquet-like formalism is  of particular interest in interpreting the appearance of Floquet-type sidebands.

To this end, a generalisation of the Floquet formalism for varying pulse amplitudes is presented in \sref{sub:ttpMod}. This formalism is called $t-t^\prime$ due to the separation of two time scales, one related to the evolution of the envelope function and the other to the periodic oscillations. The $t-t^{\prime}$ formalism~\cite{Howland1974,Peskin1993,Grifoni1998,Drese1999, Holthaus2015} has recently been applied to the case of strong pulses in two-level systems~\cite{Ikeda2022}. In contrast to this work, here we are concerned with its application to a momentum-dependent Hamiltonian with the aim of constructing the time-evolved Floquet Bloch spectrum. Before introducing the Floquet formalism for non-periodic drivings, the standard Floquet formalism is briefly reviewed in \sref{sub:FQPeriodic} in order to set the notation and to define a term of comparison with the later extension to the $t-t^{\prime}$ formalism.


\subsection{Floquet formalism for time-periodic drivings} \label{sub:FQPeriodic}

Floquet theory for time-periodic systems is used to deal with Hamiltonians of the form
\begin{equation} \label{eq:FQP:Ht}
    \ham_\mathrm{per}(t) = \ham_0 + V( t)~,
\end{equation}
where $\ham_0$ is the time-independent Hamiltonian of the system and $V(t)$ describes the coupling with the periodic driving such that $V(t) = V(t+T)$. Here, the period $T$ is related to the driving frequency, {\em i.e.}, $\omega=2\pi/T$.
%
%
The final goal is to solve the time-dependent Schrödinger equation (TDSE)
\begin{equation} \label{eq:FQP:TDSE}
    i \hbar \frac{d}{dt}~\ket{\psi(t)} = \ham_\mathrm{per}(t)~\ket{\psi(t)}~,
\end{equation}
for the state $\ket{\psi(t)}$. Due to the time periodicity of $V(t)$, the TDSE is solved in a convenient basis, called Floquet basis $\ket{\phi_b^F(t)}$, where $b$ is the band index. The state is expanded in the Floquet basis~\cite{Shirley1965}
\begin{subequations}
\begin{equation} \label{eq:FQP:linearcombF}
   \ket{\psi(t)} = \sum_b f_b~\ket{\phi_b^F(t)}~,
\end{equation}
where $f_b$ are complex constants, the summation is carried over the bands and the Floquet functions are defined as
\begin{equation} \label{eq:FQP:phiF}
   \ket{\phi_b^F(t)}= e^{-i\xi_b t/\hbar}~\ket{u_b(t)}~.
\end{equation}
\end{subequations}
Here $u_b(t+T)=u_b(t)$ is a periodic function, while the exponential part is given by the so-called Floquet \textit{quasi-energy} $\xi_b$. Note that the energy is not a conserved quantity of the system due to the breaking of time reversal invariance. However, the periodicity creates a discrete invariance, which translates into a conservation of the quasi-energy modulo the driving frequency. All Floquet solutions to the TDSE can be shifted to quasi-energies that fall within the same interval of width $\hbar \omega$. This leads to the definition of the \textit{Floquet-Brillouin zone} (FBZ) such that the first FBZ (1FBZ) contains all the quasi-energies in the interval $ -\hbar \omega /2  < \xi^\mathrm{1FBZ} <  \hbar \omega/2$. The energies $\xi^\mathrm{1FBZ}$ shifted by integer values of $\hbar \omega$ form the so-called Floquet \textit{replicas} or \textit{sidebands}
\begin{equation}
    \xi_{(b,l)} = \xi_b^{\mathrm{1FBZ}} + l \hbar \omega~,
\end{equation}
with $l$ an arbitrary integer.

The solution of the TDSE~\eref{eq:FQP:TDSE} for a periodic driving is then reduced to set the Floquet states~$\ket{u_b(t)}$, the quasi-energies $\xi_b$ and the projection $f_b$ of the initial states over the Floquet states. A common strategy for this is to exploit the periodic properties of $\ket{u_b(t)}$ and perform a discrete Fourier series decomposition in terms of the harmonics of the driving frequency, as explained in more detail in the  \ref{App:FQP:Expansion}.
Within the Fourier decomposition, it is possible to obtain an effective time-independent Floquet-Fourier Hamiltonian given by~\cite{Giovannini2019}
\begin{equation} \label{eq:FQP:H_FF}
    H_{mn} =
    \frac{1}{T} \int_T \dt \ham_\mathrm{per}(t) e^{i(m-n)\omega t}
    - m \hbar \omega \delta_{m,n}~.
\end{equation}
%
In the former expressions $m, n$ are integers corresponding to the \textit{harmonic} indices of the Fourier expansion. The Floquet states and the quasi-energies are then obtained by solving the time-independent eigenvalue problem in Fourier space given by \eref{eq:FQP:Heff}.

Finally, some remarks about the projection coefficients $f_b$ in \eref{eq:FQP:linearcombF} are due. In the case of the time-periodic driving, the coefficients $f_b$ are time-independent. They are defined as the projection of the states onto the Floquet functions for any time $t$ as
\begin{equation} \label{eq:FQP:f_b}
    f_b = \braket{\phi^F_b (t)}{\psi(t)}~.
\end{equation}
In particular, given an initial state, the coefficients $f_b$ can be calculated at $t=0$. Then, for a periodic driving, the time evolution is dictated by \eref{eq:FQP:linearcombF} and the dynamics is encoded in the basis itself, so that the coefficients $f_b$ remain constant.

The projection coefficients $f_b$ are of particular interest because they can be related to the experimental imaging of the Floquet spectra. In particular, in the case of periodic Floquet driving and neglecting the effect of the probe pulse in photoemission spectroscopy experiments, the transition probability amplitude is expected to be proportional to~\cite{Giovannini2019}
\begin{equation} \label{eq:FQP:P}
    P(\Omega)
    = \sum_{b, m} |f_b|^2 |f_{m,b}|^2 ~ \delta (\xi_b/\hbar +m\omega-\Omega)~,
\end{equation}
where the $f_{m,b} \equiv \langle u_{b}^{(m)}\mid \psi(0)\rangle$ is defined as the projection of the initial state onto the $m$-th replica (see ~\ref{App:P} for further details). In view of this expression, $|f_{m,b}|^2$ can be interpreted as the occupations of the $m$-th FBZ, and the intensity of the photoelectron spectroscopy signal is expected to be proportional to this magnitude.

\subsection{Floquet \texorpdfstring{$t-t^{\prime}$}\ \ formalism} \label{sub:ttpMod}

Within the Floquet $t-t^{\prime}$ formalism one decouples the amplitude of the driving from the oscillatory part of the pulse, which is still considered to be time-translationally invariant.
The Hamiltonian of the driven system is then expressed as
\begin{equation} \label{eq:FQttp:Ht}
    \ham_{\mathrm{pulse}}(t) = \ham_0 + a(t) V( t)~.
\end{equation}
Here $V(t) = V(t+T)$ is periodic and $a(t)$ describes the amplitude envelope. The driving frequency $\omega$ is considered to be constant and the only main assumption is that a factorisation of the driving into an envelope and a fast oscillation are appropriate for the non-periodic pulse.

\par If the evolution of the envelope is considered separately from the time-periodic part, the TDSE for the Hamiltonian~\eref{eq:FQttp:Ht} can be expressed as
\begin{equation} \label{eq:FQttp:TDSE}
    i \hbar \frac{d}{dt}~\ket{\psi(t)} = \ham_{\mathrm{pulse}}(a(t), t)~\ket{\psi(t)}~,
\end{equation}
where $\ham_{\mathrm{pulse}}$
has a parametric dependence on the amplitude due to the factorisation of the vector potential. For a fixed amplitude $a(t)$, equation~\eref{eq:FQttp:TDSE} is the same as that for a periodic Floquet driving, whose solution can be expanded in the \textit{instantaneous} Floquet basis~\eref{eq:FQP:linearcombF}:
\begin{equation} \label{eq:FQttp:linearcombF}
   \ket{\psi(t)} = \sum_b f_b(t)~\ket{\phi_b^F(a,t)}~.
\end{equation}
Although \eref{eq:FQttp:linearcombF} is similar to \eref{eq:FQP:linearcombF}, in the non-periodic case the expansion coefficients are time-dependent and the Floquet basis depends on the amplitude $a(t)$. The instantaneous Floquet states involved in the former expansion are given, as in the periodic Floquet formalism, by a periodic function times a phase factor related to the quasi-energy by
\begin{equation} \label{eq:FQttp:phiF}
   \ket{\phi_b^F(t)}= e^{-i\xi_b (a) t/\hbar}~\ket{u_b(a, t)}~.
\end{equation}
Substituting this factorisation into \eref{eq:FQttp:linearcombF}, we obtain the following expression for the solution of the TDSE with parametric dependence on the amplitude:
\begin{equation}
   \ket{\psi(t)} = \sum_b f_b(t)~e^{-i\xi_b (a)~t/\hbar}~\ket{u_b(a, t)}~.
\end{equation}
Finally, the following quantities related to the Fourier expansion are defined
\begin{equation} \label{eq:FQttp:ut_bl}
   \ket{u_{\alpha}(a,t)}~\equiv e^{i l \omega t} ~\ket{u_b(a,t)}~,
\end{equation}
where $\alpha \equiv (b,l)$ labels the band and the harmonics. The quasi-energy is also related to its replicas by
\begin{equation}\label{eq:FQttp:xi_bl}
   \xi_{\alpha}(a)~\equiv  \xi_b(a) + l \hbar \omega~.
\end{equation}
Using the former harmonic decomposition, the evolution of the states of $\ham_\mathrm{pulse}$ as a function of the instantaneous Floquet states is then written as
\begin{equation} \label{eq:FQttp:psit_dec}
   \ket{\psi(t)} = \sum_{\alpha} c_{\alpha} (t)~\ket{u_{\alpha}(a(t), t)}~,
\end{equation}
where $c_{\alpha}(t)$ are the expansion coefficients with respect to the harmonics. Although $f_b$ and $c_{\alpha}$ obey a similar expression, $c_{\alpha} (t)$ includes by definition the contribution of the (instantaneous) Floquet quasi-energy. Therefore, a direct identification of the two quantities is not possible. However, using \eref{eq:FQttp:psit_dec} as an \textit{ansatz} for the TDSE, the following expression 
\begin{equation} \label{eq:FQttp:c_t}
    i \hbar \frac{d c_{\alpha}}{dt} = \sum_{\beta} \ham^{tt^\prime}_{\alpha \beta}(a(t))~c_\beta(t)~,
\end{equation}
for the evolution of the expansion coefficients is obtained,
where the Hamiltonian $\ham^{tt^\prime}_{\alpha \beta}$ is given by
\begin{subequations}
\begin{align}
    \ham^{tt^\prime}_{\alpha \beta} & \equiv \delta_{\alpha \beta} \xi_{\alpha}(a(t))
    - i \frac{da}{dt} \c{G}^{tt^\prime}_{\alpha \beta} (a(t))~,
    \label{eq:FQttp:Cham}\\
    \c{G}^{tt^\prime}_{\alpha \beta} (a(t)) & \equiv  \int_{0}^T \frac{dt'}{T} \braket{u_\alpha(a(t),t')|\partial_a}{u_\beta(a(t),t')}~.
    \label{eq:FQttp:Gab}
\end{align}
\end{subequations}
Here, $\partial_a = \partial/\partial a$. The two contributions in expression~\eref{eq:FQttp:Cham} account for the phase acquisition associated with the quasi-energies and the transition between Floquet replicas given by the $\c{G}^{tt^\prime}_{\alpha \beta}$ term.

\par Note that the instantaneous Floquet states in \eref{eq:FQttp:Gab} are assumed to be differentiable.
In order to satisfy this condition and to avoid a spurious  phase, the overall parallel transport is required to fullfil
\begin{equation} \label{eq:FQttp:gauge}
    \braket{u_\alpha(a,t)|\partial_a}{u_\alpha(a,t)}=0~.
\end{equation}
This condition must be imposed on the states before the time evolution of $c_\alpha(t)$ can be calculated according to \eref{eq:FQttp:c_t}.
A minimal example of the $t-t^{\prime}$ formalism is analysed in \sref{sub:LinDirac} for the case of a linear Dirac Hamiltonian driven by a linearly polarised pulse.


\subsection{Adaptation of the Floquet \texorpdfstring{$t-t^{\prime}$}\ \ formalism to Floquet-Bloch Hamiltonians} \label{sub:ttpBloch}

So far, the $t-t^{\prime}$ formalism has been derived in a very general and schematic way.
Our main aim, however, is to apply this formalism to the calculation of Floquet bands for a non-periodic driving applied to a spatially translation invariant Hamiltonian in $D$ dimensions, which includes $N$ bands.
In the absence of the external driving, the Hamiltonian of this system is given by a Bloch Hamiltonian $\widehat{H}_0({\bm k})$ parametrically dependent on the momentum ${\bm k} = (k_1, \ldots ,  k_D)$.
The external driving is included via minimal coupling, i.e. by replacing $\hbar {\bm k} \to e {\bm A}(t) + \hbar {\bm k}$, where ${\bm A}(t)$ is the vector potential of the electromagnetic field.
The time-dependent part of the Hamiltonian~\eref{eq:FQttp:Ht} is then given by
\begin{equation} \label{eq:FQk:W_t}
    \widehat{W}({\bm k}, t) = \widehat{H}_0\left[ {\bm k}+e {\bm A}(t)/\hbar \right] - \widehat{H}_0({\bm k})~,
\end{equation}
where $\widehat{W}({\bm k}, t)$ is an operator that can be represented by a $N\times N$ matrix that inherits the non-periodicity in time of the vector potential under consideration.

\par In order to directly apply the formalism presented earlier, the time-dependent part should be factorised to satisfy \eref{eq:FQttp:Ht}.
This means
\begin{equation} \label{eq:FQk:Wfactor}
 \widehat{W}(t)=a(t)\widehat{V}(t)~,
\end{equation}
with $a(t)$ a non-periodic scalar function and $\widehat{V}(t)$ a periodic operator, i.e. $\widehat{V}(t) = \widehat{V}(t+T)$. However, this factorisation is not generically possible in the case of a Hamiltonian with complicated dependencies on the momenta, as is usually the case for ${\bm k}\cdot {\bm p}$ low-energy Hamiltonians. For example, different polynomial orders in the momenta make this task much more difficult.

\par To treat these Hamiltonians in the $t-t^{\prime}$ formalism, we define the auxiliary variable $\eta(t)$ so that the Hamiltonian can be written generally as~\cite{Drese1999}
\begin{equation} \label{eq:FQk:Http}
    \widehat{H}({\bm k}, \eta(t), t) = \widehat{H}_0({\bm k}) + \widehat{W}({\bm k}, \eta(t), t)~,
\end{equation}
with $\widehat{W}({\bm k}, \eta, t+T) = \widehat{W}({\bm k}, \eta, t)$ for a fixed $\eta$.
Thus, for each fixed value of $\eta$, the corresponding Floquet basis is defined by
\begin{align} \label{eq:FQk:phi_F}
    \ket{\phi_b^F({\bm k}, \eta, t)}= e^{-i\xi_b ({\bm k}, \eta) t/\hbar}~\ket{u_b({\bm k}, \eta, t)}~.
\end{align}
The Floquet quasi-energies $\xi_b ({\bm k}, \eta)$ and the Floquet spinors $\ket{u_b({\bm k}, \eta, t)}$ can be calculated employing the Floquet-Fourier expansion in \eref{eq:FQP:H_FF} upon substitution of $\ham_\mathrm{per}(t) \to \widehat{H}({\bm k}, \eta, t)$. Note that this is possible because, even if the periodic part is not easily factorisable, the frequency of the period $T$ is well defined at fixed $\eta$ and it is the same for all $\eta$ values. Thus, the $\eta$-dependent Fourier modes are defined by \eref{eq:FQP:u_bm} as
\begin{equation} \label{eq:FQk:u_bm}
    \ket{u_b ({\bm k}, \eta, t)} = \sum_{m= -\infty}^\infty e^{-im\omega t} ~\ket{u_b^{(m)}({\bm k}, \eta)}~.
\end{equation}
The Floquet replicas are obtained by generalising ~\eref{eq:FQttp:ut_bl} and \eref{eq:FQttp:xi_bl} where, as before, the double index $\alpha \equiv (b,l)$ indicates the band $b$ and the replica $l$:
\begin{align}
    \ket{u_{\alpha}({\bm k}, \eta,t)} &~\equiv e^{i l \omega t} ~\ket{u_b({\bm k}, \eta,t)}~,\\
    \xi_{\alpha}({\bm k}, \eta) & ~\equiv  \xi_b({\bm k}, \eta) + l \hbar \omega~.
\end{align}
Although the definition of the replica index is arbitrary, in the specific case of Floquet-Bloch Hamiltonians it is particularly convenient to define the states in such a way that, in the limit without external driving, the replica with $l=0$ coincide with the original band dispersion. In other words, instead of using the 1FBZ, the energy $\xi_b({\bm k}, \eta)$ is defined by the limit
\begin{equation} \label{eq:FQk:limEk}
  \lim_{\widehat{W}({\bm k}, t)\to 0}  \xi_b({\bm k}, \eta(t)) \to \epsilon_b ({\bm k})~,
\end{equation}
where $\epsilon_b ({\bm k})$ defines the $b$-th band of the unperturbed Hamiltonian from the time-independent Schr\"{o}dinger equation $\widehat{H}_0({\bm k}) \psi_b = \epsilon_b ({\bm k}) \psi_b $.

\par Finally, the TDSE for the Hamiltonian $\widehat{H}({\bm k}, \eta, t)$ can be written employing the Floquet decomposition
\begin{align} \label{eq:FQk:psit_dec}
    \ket{\psi(t)} = \sum_{\alpha} c_{\alpha} ({\bm k}, t)~\ket{u_{\alpha}({\bm k}, \eta(t), t)}~,
\end{align}
which leads to the form, equivalent to \eref{eq:FQttp:c_t},
\begin{equation} \label{eq:FQk:c_t}
i \hbar \frac{d c_{\alpha}({\bm k}, t)}{dt} = \sum_{\beta} \widehat{H}^{tt^\prime}_{\alpha \beta}({\bm k}, \eta(t))~c_\beta({\bm k}, t)~,
\end{equation}
where
\begin{subequations}
\begin{align}
    \widehat{H}^{tt^\prime}_{\alpha \beta}({\bm k}, \eta) & \equiv \delta_{\alpha \beta} \xi_{\alpha}({\bm k}, \eta)
    - i \frac{d\eta}{dt} \widehat{G}^{tt^\prime}_{\alpha \beta} ({\bm k}, \eta)~,
    \label{eq:FQk:Cham}\\
    \widehat{G}^{tt^\prime}_{\alpha, \beta}({\bm k},\eta)
    & \equiv \sum_{m=-\infty}^\infty \braket{u_{b}^{(m+l-l')}({\bm k}, \eta)}{\partial_\eta u_{b'}^{(m)}({\bm k}, \eta)}~.
    \label{eq:FQk:Gab}
\end{align}
\end{subequations}
The indices are defined as $\alpha \equiv {(b,l)}$ and $\beta \equiv (b',l')$, and the partial derivative is denoted by $\partial_\eta = \partial/\partial_\eta$. More details on the derivation of this expression are given in~\ref{App:eqGab}.

This formulation of the $t-t^{\prime}$ problem is particularly convenient because it does not require the explicit factorisation in \eref{eq:FQk:Wfactor} and, moreover, it gives the evolution of the expansion coefficients $c_\alpha(t)$ within a time-independent effective Hamiltonian~\eref{eq:FQk:Cham} constructed by using the static Fourier components. In addition, it is possible to associate the parameter $\eta$ with the more convenient time-dependent function in the problem and then play with the properties of the partial derivative in \eref{eq:FQk:Gab} to map the Fourier problem into the appropriate formulation for a numerical solution.
In this case, the parallel transport condition has to be implemented as a function of $\eta$ and reads
\begin{equation} \label{eq:FQk:gauge}
    \braket{u_b^{(m)}({\bm k}, \eta)}{\partial_\eta u_b^{(m)}({\bm k}, \eta)}=0~.
\end{equation}

Note that throughout this section the ${\bm k}$ momenta have been treated as parameters of the Hamiltonian and the Floquet $t-t^{\prime}$ problem has been defined separately for each ${\bm k}$ mode. However, we consider it appropriate to keep the ${\bm k}$-dependence explicit in order to define more precisely the problem of factorising the time-dependent part given by \eref{eq:FQk:Wfactor} and to discuss the more convenient definition of the replicas by the limit~\eref{eq:FQk:limEk}. In fact, only by explicitly keeping the ${\bm k}$-dependence it becomes clear that it is possible to completely decouple the time evolution and the 
generally complex momentum dependence entering through the minimal substitution by means of the auxiliary parameter $\eta$, which can be uniquely defined for all ${\bm k}$ modes of the problem.


\subsection{
Comparison with 
direct solution of the TDSE} \label{sub:ttpvsUt}

The central problem of the $t-t^\prime$ formalism is to solve the differential equations~\eref{eq:FQttp:c_t} or~\eref{eq:FQk:c_t} for the evolution of the expansion coefficients $c_\alpha(t)$. With these and the time dependent Floquet basis it is possible to obtain the time evolution of the state according to equations~\eref{eq:FQttp:psit_dec} or~\eref{eq:FQk:psit_dec}. However, the solution of $\psi(t)$ is not the central result of the $t-t^{\prime}$ formalism. In fact, the evolution of the states is more easily solved by the TDSE in its differential form~\eref{eq:FQttp:TDSE} using the non-periodic Hamiltonian~\eref{eq:FQttp:Ht}. Formally, the solution of the TDSE by direct integration of the differential equation is given by the time-evolution operator
%
\begin{equation} \label{eq:U_t}
    U(t,t_0) = \mathcal{T} \exp \left[ -\frac{i}{\hbar} \int^{t}_{t_0} \ham (t) dt \right]~,
\end{equation}
where $\mathcal{T}$ denotes the time ordering. Then, the evolution of an initial state $\ket{\psi(t_0)}$ is expressed as a function of the evolution operator by the well-known expression
\begin{equation} \label{eq:FQP:psit_U}
    \ket{\psi(t)} = U(t,t_0) \ket{\psi(t_0)}~.
\end{equation}
For a Bloch  Hamiltonian with $N$ bands, the inclusion of the instantaneous Floquet basis increases the size of the problem by the Fourier-Floquet expansion, leading to a Hilbert space of size $M N \times M N$, where $M$ is the number of harmonics considered. 
The direct solution of the TDSE can only refer to the evolution of the states themselves and, due to the lack of translational invariance in time, the energy cannot be defined. The instantaneous Floquet basis used to factorise the states provides a more convenient interpretation of the dynamics in terms of the Floquet sidebands for time-dependent driving amplitudes. Thanks to the well-defined driving frequency $\omega$, it is still possible to define the time-dependent spectrum of the quasi-energies $\xi_\alpha (t)$ and to interpret the occupancy of the replicas as $|c_\alpha(t)|^ 2$.

Finally, it is also important to underline that the numerical effort of enlarging the Hilbert space dimension up to $M N \times M N$ is still not so large in comparison to the spectrum of frequencies that can be obtained by Fourier transforming $\psi(t)$ from the TDSE. In fact, thanks to the Fourier basis written for each $\eta$, the sum over $m$ corresponding to the harmonics in \eref{eq:FQk:Gab} can be truncated by analysing the support of the eigenvectors of the Floquet-Fourier expansion, similar to what is done in the usual Floquet-Fourier calculations. 


\section{Application of the $t-t^{\prime}$ formalism to Effective Dirac systems} \label{sec:Applications}


\subsection{Linear Dirac model} \label{sub:LinDirac}

\par In this section, we study the 2D linear Dirac model as a first simple example of the implementation of the $t-t^{\prime}$ formalism and as for later comparison with the Dirac model with trigonal warping analysed in \sref{sub:Bi2Se3}. The linear Dirac model driven by Floquet-like pulses has been extensively investigated in previous theoretical studies~\cite{Farrell2016, Junk2020, DiazFernandez2019}. In particular, the effects of driving on the topology of the bands have been specifically discussed, leading, e.g., to the concept of Floquet topological insulator~\cite{Oka2009, GomezLeon2013, Kitagawa2011}. 

In this work, we will not investigate the topological nature of the bands, but will focus on the interpretation of the sidebands arising from the external driving.
The system is given by the usual linear Dirac Hamiltonian
\begin{equation} \label{eq:LinDir:H0}
    H_0 (k_x,k_y) = \hbar v \left( k_x \sy - k_y \sx \right)~.
\end{equation}
For the sake of simplicity, a general elliptic driving is considered, given by the following in-plane vector potential
\begin{equation} \label{eq:LinDir:Aelip}
    {\bm A} (t) = \left[ a_x(t) \sin (\omega t), ~a_y (t) \sin(\omega t + \theta_0), 0  \right]~,
\end{equation}
where $a_x(t)$ and $a_y(t)$ are the time-dependent amplitudes of the driving, $\omega$ is the driving frequency and $\theta_0$ is the initial phase shift between the two components. The circular driving is obtained for $\theta_0 = \pi/2$ and $a_x(t)=a_y(t)$. Linearly polarised pulses are obtained by setting one of the amplitudes to zero, e.g. $a_y(t) = 0$. By minimal substitution, the vector potential enters the Dirac Hamiltonian as
\begin{subequations} \label{eq:LinDir:H_W}
\begin{align} 
    H (k_x, k_y, t)  & =
      H_0(k_x, k_y)
    + W_x (t) + W_y (t)~, \\
    W_x (t) & = e v a_x(t) \sin(\omega t) \sy ~, \\
    W_y (t) & = - e v a_y(t) \sin(\omega t + \theta_0) \sx~,
\end{align}
\end{subequations}
where $e$ is the elementary charge.

For concreteness, we consider a Gaussian envelope
\begin{align} \label{eq:LinDir:Gaussian}
    a_i (t) & = A_i e^{-(t/\tau)^2}~,
\end{align}
where $i = x, y$, $A_i$ is the maximum amplitude and $\tau$ is a real parameter giving the width of the Gaussian pulse. The most direct identification of the parameter $\eta$ for this pulse is given by the Gaussian modulation, i.e. by $\eta(t) = e^{-(t/\tau)^2}$. Note that in the case of the Gaussian pulse exerted to the linear Dirac Hamiltonian, the factorisation given by \eref{eq:FQk:Wfactor} is straightforward due to the mere presence of linear terms in momenta in $H_0$. 

For a fixed $\eta$, the expansion over the Fourier harmonics~\eref{eq:FQP:Heff} leads to an effective Hamiltonian with the simplified structure of a monocromatic field
\begin{equation} \label{eq:LinDir:HF}
    H_F =
\begin{pmatrix}
\ddots  &   Q   & 0      &        &        \\
Q^\dagger  & H_0+\hbar \omega & Q & 0 &         \\
0  &  Q^\dagger & H_0 & Q & 0 \\
   &   0 & Q^\dagger & H_0 - \hbar \omega & Q  \\
  &        & 0 & Q^\dagger & \ddots  \\
\end{pmatrix}~,
\end{equation}
where the momentum dependence was omitted in the Floquet effective Hamiltonian $ H_F = H_F(k_x, k_y, \eta )$ as well as in the Dirac Hamiltonian $H_0 = H_0 (k_x, k_y)$.
The term that couples the replicas is
\begin{equation} \label{eq:LinDir:Qop}
    Q = -\, \frac{i}{2}\,e v A_x \eta  \sy
    + \frac{1}{2}\,e v A_y \eta [i \cos (\theta_0)- \sin(\theta_0)] \sx~.
\end{equation}

The diagonalisation of $H_F$, equation~\eref{eq:LinDir:HF}, yields the Floquet quasi-energies $\xi_b({\bm k},\eta)$ as well as the set of Fourier modes $\ket{u^{(m)}_b({\bm k}, \eta)}$ with $b= 1, 2$ for the two bands of the model. To compute the quasi-spectrum, the Fourier expansion has to be truncated. 
In this case, the simple monochromatic structure already yields a very reduced support of the Floquet vectors in the harmonic space. In fact, only replicas up to the first order are coupled by the driving and few Fourier components are needed in the expansion to obtain a reliable result. 


\subsubsection{Linearly polarized Gaussian pulse}

The case of linear polarization is particularly instructive for the interpretation of the sidebands. The spectrum for 
$k_y=0$ is not modified by the external driving and there are no gaps in the spectrum. This can be proved by checking that the commutator of the Hamiltonian without perturbation $H_0(k_x,k_y=0)$ and the correction generated by the external pulse $W_x(t)$ commute
\begin{equation}
    [\ham_0 (k_x, k_y = 0), W_x(t)] = 0~.
\end{equation}
Thus the eigenvectors of $H_0(k_x,k_y=0)$ still diagonalize the complete Hamiltonian $H(k_x,k_y=0,t)$. Therefore, for linear polarization in the $x$-direction, the TDSE can be integrated, leading for the states at $k_{y}=0$ to~\cite{Farrell2016}
\begin{equation} \label{eq:LinDir:L_psi_int}
    \ket{\psi ({k_x,k_y = 0, s})} =  e^{-is v k_x (t-t_0)} 
     e^{is ev A_x \int^t_{t_0} dt' \eta(t) \sin(\omega t')/\hbar}
    ~ \ket{\psi_0 ({k_x,k_y = 0, s})}~,
\end{equation}
where $s = \pm 1$ is the band index and $\psi_0$ denotes the eigenstates of the original Dirac cone. The driving only causes a phase change in the states.

\begin{figure}
    \centering
    \includegraphics[width=0.8\columnwidth]{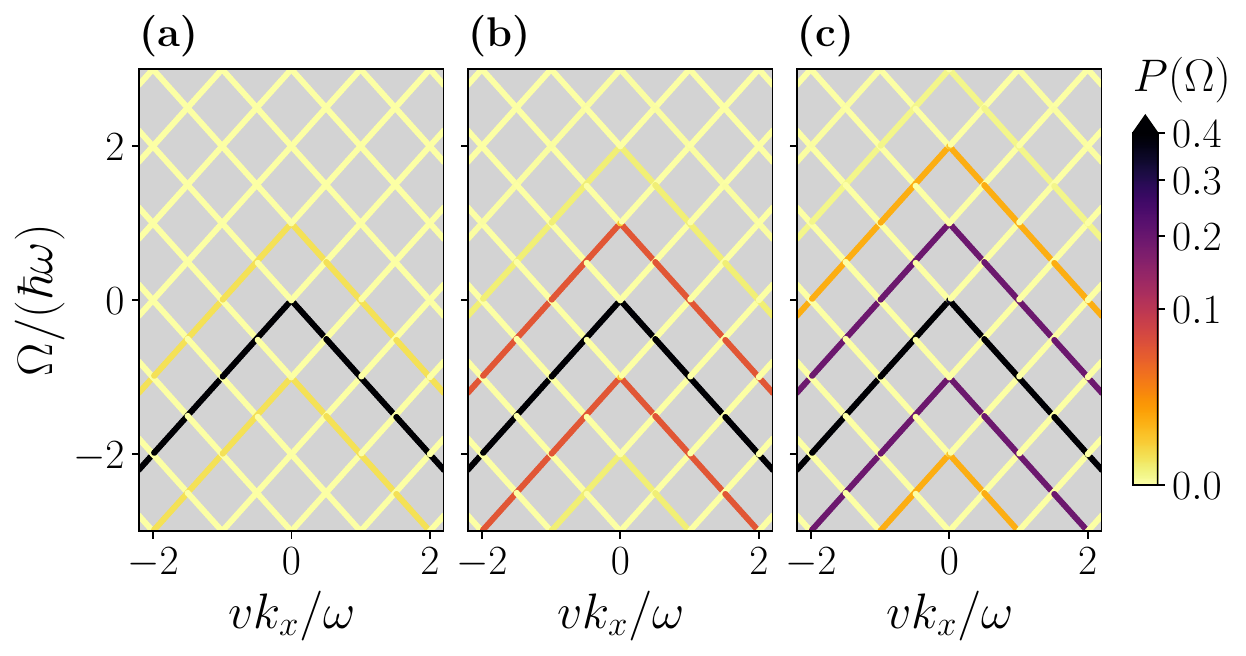}
        \caption{
        Transition probability amplitude $P(\Omega)$ (colour coded) for the linear Dirac Hamiltonian at $k_y=0$ in the case of the linearly polarized driving in $x$ (i.e. by setting $A_y = 0$). The pulse amplitudes are held fixed at $ev A_x \eta /(\hbar \omega) = \{ 0.1, 0.5, 1 \}$ in panels (a), (b) and (c), respectively. The amplitude of $P(\Omega)$ is computed according to \eref{eq:FQP:P}. The initial state $\psi_0$ employed in the calculations is the valence state for the Hamiltonian~\eref{eq:LinDir:H0} at $k_y=0$. }
        \label{fig:LinDirac:P}
\end{figure}

Even if the time evolution dictated by the direct integration of the TDSE is trivial for the state in \eref{eq:LinDir:L_psi_int}, the interpretation of the exponential factor as an equivalent energy in the corresponding Floquet picture implies a dynamics in the population of the replicas. In fact, starting from the simplest case of the static Floquet picture obtained by fixing $\eta$, the corresponding phase gained 
is related to the occupation of the Floquet replicas. The results for fixed $\eta$ are shown in \fref{fig:LinDirac:P}, where the transition probability amplitude associated with the photoemission is calculated according to expression~\eref{eq:FQP:P}. The initial state chosen is a Bloch valence band state for the system in the absence of external driving, i.e. the valence state of $H_0(k_x, k_y=0)$. The initial state in the valence branch of the Dirac cone is pumped to the nearby replicas, interpreted as photon-dressed bands, with an enhanced contribution upon increasing the pulse amplitude. Thus, in \fref{fig:LinDirac:P}(c), which corresponds to the higher amplitude studied, the state is spread over the two upper and lower replicas. In contrast, in panels (a) and (b) the density is located mainly in the valence band of the original Dirac cone. It is clear from panel (a) that in the limit of vanishing amplitude $A_x$, the obtained Floquet band structure still shows the Floquet replicas due to the finite frequency $\omega$. However, in this limit, the harmonic Fourier series becomes a purely mathematical tool and the sidebands do not describe any populated physical state. For this reason it is important to discuss the Floquet-Bloch band structure using observables such as the transition amplitude or the time-averaged density of states~\cite{Uhrig2019, Oka2009, Rudner2020, Zhou2011} to define physical quasi-energies.

In \fref{fig:LinDirac:P} the linear polarisation only populates replicas of the same band. This corresponds to a non-zero transition probability between replicas from the same band which is consistent with the form of the coupling $Q$ in \eref{eq:LinDir:Qop} for $A_y=0$. Thus, in \fref{fig:LinDirac:P}, the initial valence state spreads only over valence replicas. A mixed initial state would have produced occupancies in the conduction replicas as well, but in this case both types of dynamics, valence to valence sidebands and conduction to conduction sidebands, are decoupled.

\begin{figure}
        \centering
        \includegraphics[width=0.6\columnwidth]{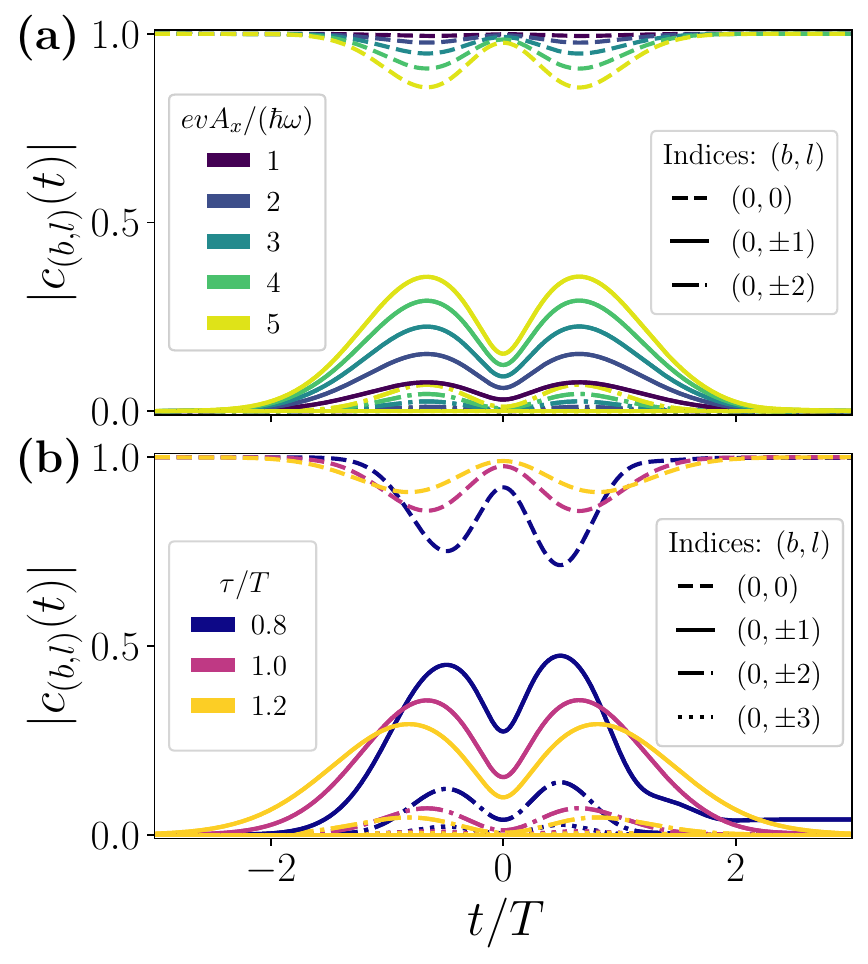}
        \caption{Absolute value of the expansion coefficients $c_\alpha(t)$ as a function of time for the linearly polarised driving from the solution of \eref{eq:FQk:c_t} for $v k_x/\omega = 0.1$ and $k_y=0$.
        In panel (a), the $|c_\alpha(t)|$ are plotted for fixed width $\tau/T = 1$ and varying the maximum amplitudes $ev A_x/(\hbar \omega) = \{1, 2, 3, 4, 5 \}$, in the colour code.
        The line type corresponds to the order of the replicas, as indicated in the legend.
        The notation between brackets is $(b,l)$, where $b$ is the band index and $l$ is the replica index.
        In panel (b), the $|c_\alpha(t)|$ are plotted for fixed $ev A_x/(\hbar \omega) = 5$ and $\tau/T = \{ 0.8, 1, 1.2\}$, as indicated in the colour code.
        The initial state considered is the valence eigenstate of $H_0(k_x, k_y=0)$, i.e. the state in band $b=0$.  }
        \label{fig:LinDirac:Lcalpha}
\end{figure}

Next, we include the time dependence in the parameter $\eta$ to solve the expansion coefficients $c_\alpha(t)$ from \eref{eq:FQk:c_t}. For sake of concreteness we fix $vk_x/\omega =0.1$ and compute the evolution of the expansion coefficients as a function of time. The results are shown in \fref{fig:LinDirac:Lcalpha} for different maximum amplitudes $A_x$ and by varying the width of the pulse $\tau$, in panels (a) and (b), respectively. The initial state considered is the valence eigenstate of the unperturbed Hamiltonian $H_0(k_x, k_y=0)$. In double-index notation this state corresponds to $(b,l)= (0,0)$ and the initial occupation is expressed by having only a non-zero expansion coefficient $c_{(0,0)} = 1$ for $t \to -\infty$ . Due to the external Gaussian pulse, the initial state $(0,0)$ is depleted and the sidebands are occupied, as can be seen in panels (a) and (b) for different pulse strengths and widths. Since only the same band replicas are coupled by the external pulses, only the expansion coefficients of the valence bands, i.e. $b=0$, are non-zero. In particular, for the parameters chosen in panels (a) and (b), only replicas up to 2nd and 3rd order, respectively, are populated.

In \fref{fig:LinDirac:Lcalpha}(a) the effect of the pulse is studied for different amplitudes. The depletion of the original band towards the sidebands is stronger for increasing maximum pulse amplitude $A_{x}$, consistent with the linear increase of the coupling $Q$ with $A_x$ in \eref{eq:LinDir:Qop}. Note that the maximum spreading of the occupancy towards different replicas is achieved when the derivative of the pulse envelope is maximum, i.e. at $t = \pm \tau/\sqrt{2}$.
Figure~\ref{fig:LinDirac:Lcalpha}(b) shows the effect of pulse width $\tau$. In this case, increasing $\tau$ decreases the transition rate to higher replicas. This can be understood from \eref{eq:FQk:c_t}: the term $\widehat{G}({\bm k}, \eta)$ does not depend on the pulse width, while the prefactor of the derivative $d \eta / dt$ is indeed increased for shorter pulses, leading to a stronger effect of the pulse by decreasing $\tau$. Therefore, for pulses that reach the same maximum value of the amplitude, the strongest coupling between replicas is achieved by increasing the sharpness of the pulse envelope. Thus, the increase in the derivative of the pulse envelope is related to a larger change in $c_\alpha(t)$ for a given $\alpha$. Conversely, in the limit of an infinitely slow envelope variation, the derivative $d \eta / dt$ tends to zero and the expansion coefficients are simply given by
\begin{equation}
    c_\alpha (t) = c_\alpha(t=0)~e^{-i\xi_b t/\hbar}~.
\end{equation}
Thus, in this adiabatic limit, $|c_\alpha|$ remains constant. 

\begin{figure}
        \centering
        \includegraphics[width=0.8\columnwidth]{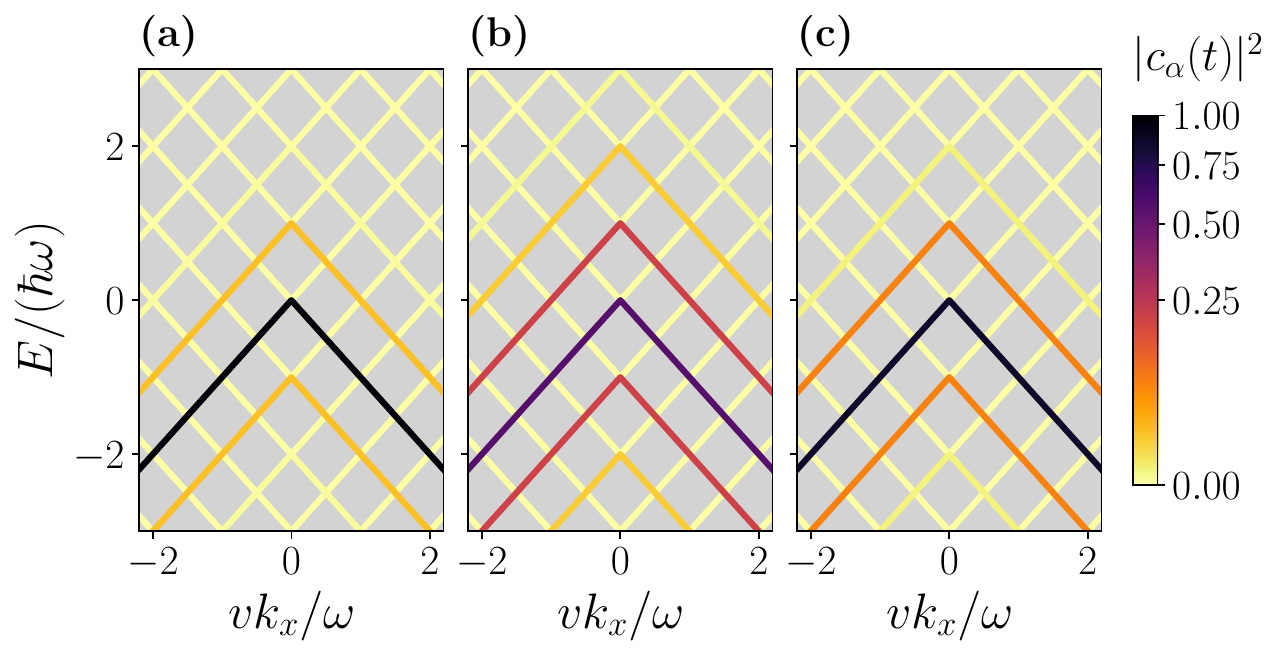}
        \caption{Expansion coefficients $|c_\alpha(t)|^2$ projected over the Floquet spectrum for three times $t/T = \{- 1.2, -0.48, 0\}$, in panels (a), (b) and (c), respectively.
        The pulse considered is a linearly polarised pulse with $\tau/T = 0.8$ and maximum amplitude $ev A_x/(\hbar \omega) = 5$.
        The initial state considered is the valence eigenstate of $H_0(k_x, k_y=0)$.
        The spectrum is plotted for $k_y=0$.}
        \label{fig:LinDirac:Lspectrum}
        \end{figure}

Finally, note that the coupling between Floquet replicas given by \eref{eq:LinDir:Qop} is independent of the momenta, namely the dynamics of all $k_x$ states with $k_y=0$ is equivalent up to an initial phase shift. This is consistent with the direct integration of the TDSE in the expression~\eref{eq:LinDir:L_psi_int} and can also be demonstrated by showing that the eigenvectors of $H_0(k_x, k_y=0)$ are the eigenstates of $\sigma_y$ and hence equation~\eref{eq:FQk:c_t} for different $k_x$ differs only by the term $\xi_\alpha$ which gives the phase acquisition. In conclusion, the results in \fref{fig:LinDirac:Lcalpha} are actually valid for any $k_x$ mode. Given this, it is almost straightforward to construct a snapshot of the evolution of the occupation of the Floquet band structure as a function of time. This is shown in \fref{fig:LinDirac:Lspectrum}, where the squared modulus of the amplitude, $|c_\alpha(t)|^2$, is projected onto the Floquet spectrum for three time instants corresponding to $\eta(t) = \{0.1, 0.7, 1\}$, giving $t/T = \{-1.2, -0.48, 0\}$. From \eref{eq:FQk:psit_dec} it is clear that $|c_\alpha(t)|^2$ can indeed be interpreted as the time-dependent occupancy of the different Floquet replicas. Thus, \fref{fig:LinDirac:Lspectrum} can be undestood as the time evolution of the Floquet-Bloch band occupancies.


\subsubsection{Circularly polarized Gaussian pulse}

In the case of circular polarisation, the operator describing the time-periodic field and the original Hamiltonian do not commute, in contrast to linear polarisation~\cite{Farrell2016}. The driving then produces not only a rigid shift of the bands, but also a hybridisation of the Floquet replicas, leading to gaps in the spectrum. Avoided crossings occur at zero energy and at resonances where the original bands were separated by multiples of $\hbar \omega$~\cite{Junk2020}. 

\begin{figure}
    \centering
    \includegraphics[width=0.8\columnwidth]{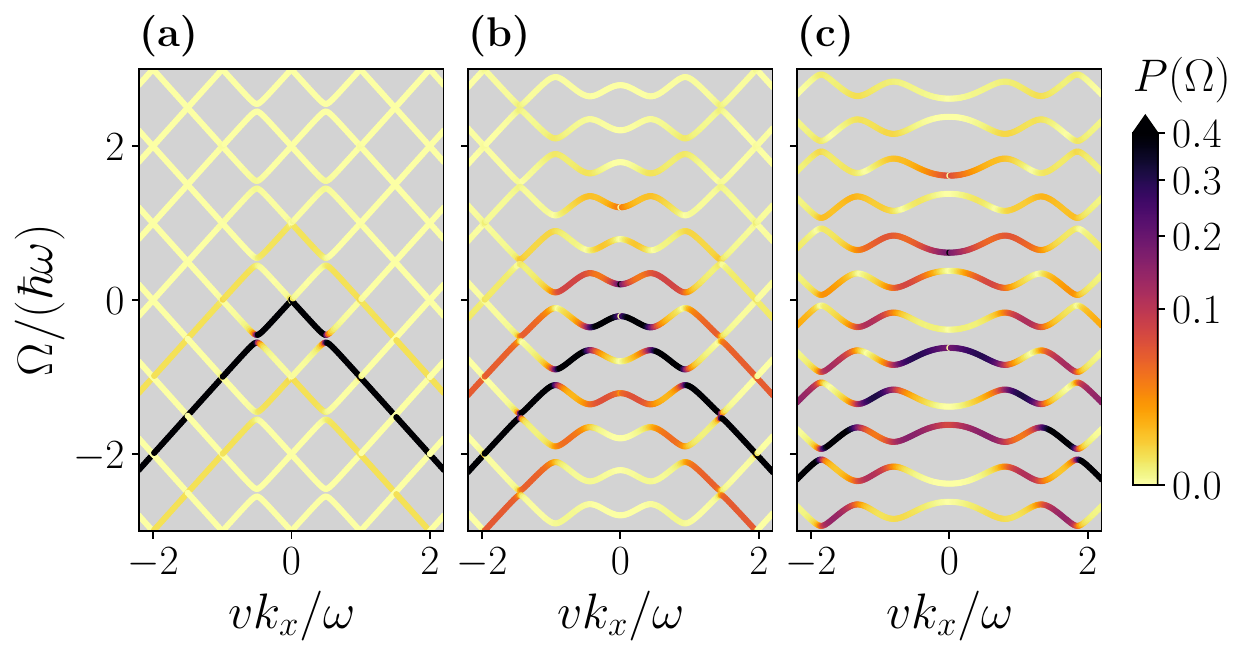}
        \caption{Transition probability amplitude for the Dirac Floquet spectrum for a circularly polarized driving at $k_y=0$ (i.e. by setting $A_x = A_y$ and $\theta_0 = \pi/2$).
        The pulse amplitudes are $ev A_i \eta /(\hbar \omega) = \{0.1, 0.5, 1\}$ with $i = x, y$ in panels (a), (b) and (c), respectively.
        The colour code corresponds to the amplitude of the $P(\Omega)$, calculated according to \eref{eq:FQP:P}.
        The initial state $\psi_0$ is the valence state of $H_0(k_x, k_y=0)$. }
        \label{fig:LinDirac:CPomega}
\end{figure}

The results for fixed $\eta$ are shown in \fref{fig:LinDirac:CPomega}, where the transition probability amplitudes associated with photo emission~\eref{eq:FQP:P} are projected over the bands. Again, the initial state considered is the valence band state of the Dirac cone for $H_0(k_x, k_y=0)$. Regardless of the polarisation, this initial state is pumped to the nearby replicas with an increasing efficiency depending on the pulse amplitude. In the case of circular polarisation, opposite bands are coupled allowing for transitions between their replicas, as can be seen in \fref{fig:LinDirac:CPomega}(b) and~(c) in the occupation of the replicas of the conduction band for $k_x \simeq 0$.

\begin{figure}
        \centering
        \includegraphics[width=0.8\columnwidth]{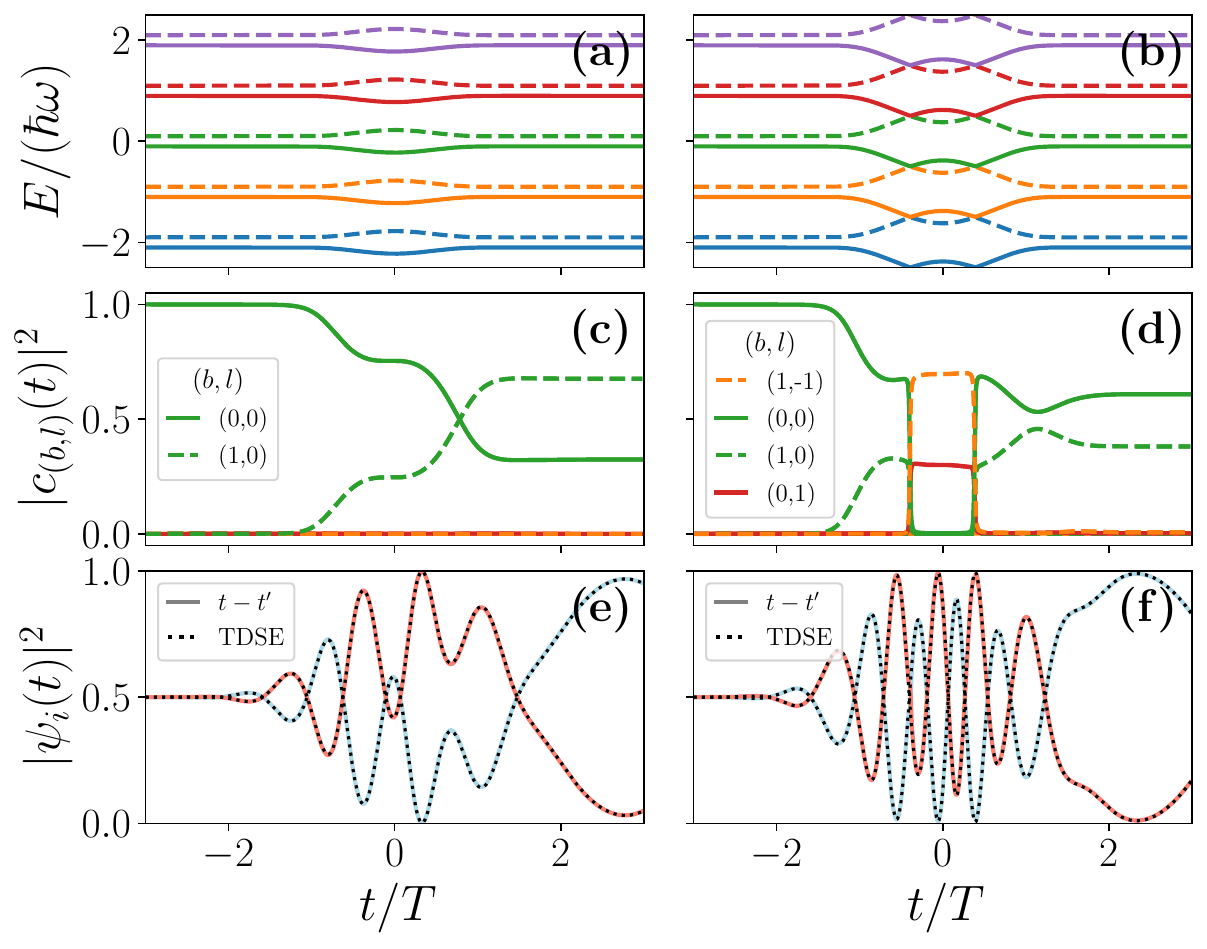}
        \caption{Floquet quasi-energies (a,b), expansion coefficients (c,d) and evolution of the wavefunction components (e,f) as a function of time for two parameter sets.
        In the left panels the maximum amplitude of the Gaussian envelope is $ev A_i/(\hbar \omega) = 0.5$ with $i = x, y$, while in the right panels $ev A_i/(\hbar \omega) = 1$.
        The other parameters are fixed to $v k_x/\omega = 0.1$, $k_y=0$ and $\tau/T = 3$.
        The notation between the brackets is $(b,l)$, where $b$ is the band index and $l$ is the replica index.
        }
        \label{fig:LinDirac:Ccalpha}
\end{figure}

Next, similar to the previous section, we account for the time dependence in the parameter $\eta$ when computing the evolution of the expansion coefficients $c_\alpha(t)$ according to \eref{eq:FQk:c_t}. For concreteness, we consider a mode near the Dirac point by fixing $vk_x/\omega =0.1$. The expansion coefficients as a function of time are plotted in \fref{fig:LinDirac:Ccalpha} for two representative parameter sets. The initial state considered for the time evolution is again a valence band state for the unperturbed Hamiltonian, i.e. $(b,l) = (0,0)$.
In the left panels of \fref{fig:LinDirac:Ccalpha} the evolution is calculated for a maximum amplitude $ev A_i/(\hbar \omega) = 0.5$ with $i = x, y$. Note that the external pulse creates a repulsion between the energy levels, see panel (a), which shifts the occupation towards the conduction band level, i.e. from $(b,l) = (0,0)$ to $(b,l) = (1,0)$. Due to the external driving, part of the valence band occupation remains in the conduction band after the pulse, see panel (c).

A stronger pulse is considered in the right panels of \fref{fig:LinDirac:Ccalpha}, corresponding to an amplitude of $ev A_i/(\hbar \omega) =1$ 
In this case, the level repulsion induced by the pulse actually couples the replicas of opposite bands, producing small gaps between them, almost imperceptible on the scale of panel~(b). This coupling leads to a shift of the occupation between the replicas as a function of time, as shown in panel~(d). In \fref{fig:LinDirac:Ccalpha}, panels (e) and (f) represent the spinorial components of $\psi(t)$ from the $t-t^{\prime}$ solution~\eref{eq:FQk:psit_dec} and the direct integration of the TDSE. The comparison of the two is a valuable check of the numerical results, which  are in perfect agreement.

\begin{figure}[htb]
    \centering
    \includegraphics[width=0.8\columnwidth]{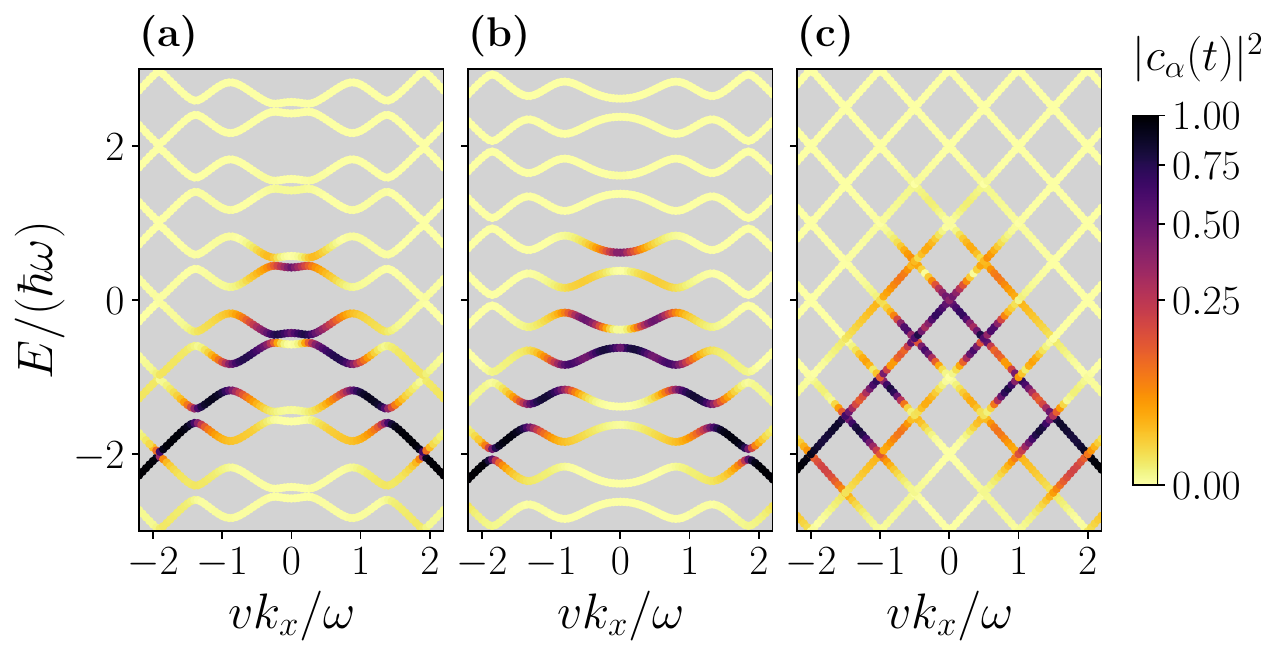}
    \caption{Expansion coefficients $c_\alpha(t)$ projected over the Floquet spectrum for three times $t/T = \{- 0.5, 0, 2\}$ in panels (a), (b) and (c), respectively, for a circularly polarized pulse with $\tau/T = 1$ and maximum amplitude $ev A_x/(\hbar \omega) = 1$.
    The initial state is the valence eigenstate of $H_0(k_x, k_y=0)$.
    The spectrum is plotted for $k_y=0$.}
    \label{fig:LinDirac:Cspectrum}
    \end{figure}

Finally, we study the effect of the circularly polarised pulse on the Floquet-Bloch spectrum: \fref{fig:LinDirac:Cspectrum} shows the projection of the expansion coefficients of the $t-t^{\prime}$ basis over the quasi-energies for three instants of the time evolution corresponding to $t/T = \{- 0.5, 0, 1\}$. The coupling between the two bands allows for transitions between valence and conduction replicas. This is already visible in panel (a) at $k_x\simeq 0$ and at $vk_x = n \omega /2$, with $n\in \mathbb{Z}$, the resonance condition for the appearance of the pulse-induced gaps. The coupling not only opens the gaps and pushes the occupation towards the replicas, but also allows for a residual occupation at the pulse end, leading to a distinctly different reconfiguration of the occupation of the Dirac cone bands [see panel~(c)]. The supplementary \fref{fig:LinDirac:Cspectrum_6panels} in \ref{App:ExtraFigs} shows additional time snapshots for completeness.


\subsection{Dirac Hamiltonian for Bi\textsubscript{2}Se\textsubscript{3} surface states} \label{sub:Bi2Se3}

In this section, the $t-t^{\prime}$ formalism is applied to study the surface states of Bi$_2$Se$_3$ in the presence of a Gaussian pulse. For this material the effective Hamiltonian of the surface states is given by a linear Dirac cone and an additional trigonal warping term~\cite{Liu2010, Shan2010, Fu2009} leading to
\begin{equation} \label{eq:Bi2Se3:H0}
    H_0 (k_x, k_y) = (c_0 + c_2\hbar^2{\bm k}^2)\mathbb{1}_2 + \hbar v (k_y\sx-k_x\sy)
    + \hbar^3 r (k_{+}^3+ k_{-}^3) \sz~,
\end{equation}
where ${\bm k}^2= k_x^2+k_y^2$ and $k_{\pm}=k_x \pm i k_y$. The system parameters are related to the bulk spectrum parameters given in Table~\ref{tab:Bi2Se3} by~\cite{Liu2010}
\begin{equation}
    c_0 = C_0 + \alpha_3 M_0~,\quad
    c_2 = C_2+ \alpha_3 M_2~, \quad
    v = v_0 \alpha_1~, \quad
    r = R_1 \alpha_1/2~,
\end{equation}
where $\alpha_1 = 0.99$ and $\alpha_3 = -0.15$ are defined to match the experimental values of the velocity and the position of the Dirac points~\cite{Xia2009} following the fit of \cite{Liu2010}.

\begin{table}[htb]
\caption{Values for the parameters of \ce{Bi_2Se_3} from \cite{Liu2010}. 
    \label{tab:Bi2Se3}} 

\begin{indented}
\lineup
\item[]\begin{tabular}{@{}*{3}{l}}
\br                              
\centre{3}{Numerical values for the parameters of Bi$_2$Se$_3$} \cr 
\mr
$\hbar v_0 = \SI{3.33}{\eV \angstrom}$  &
        $C_0 = \SI{-0.0083}{\eV}$  & 
        $M_0 = \SI{-0.28}{\eV} $  \cr 
        $\hbar^3 R_1$ = \SI{50.6}{\eV \angstrom^3} & 
        $\hbar^2 C_2 = \SI{30.4}{\eV \angstrom^2} $ &
         $\hbar^2 M_2 = \SI{44.5}{\eV \angstrom^2}$ \cr 
\br
\end{tabular}
\end{indented}
\end{table}


By including the external pulse through minimal coupling, the Hamiltonian can be written as
\begin{equation} \label{eq:Bi2Se3:Ht}
H (k_x, k_y, t)   = \ham_0 + W_x (t) + W_y (t) + W_{xy}(t)~,
\end{equation}
whith the time-dependent potentials
\begin{subequations} \label{eq:Bi2Se3:Ws}
\begin{align}
    W_x (t) & = 2 r \sz a_x^3(t)  + (c_2 \id +6 r k_x \sz) a_x^2(t)  \nonumber \\
             & ~ + [6 r (k_x^2-ky^2) \sz+ 2 c_2 k_x \id - v \sy] a_x(t)~,\label{eq:FQTrig:Vx} \\
    W_y (t) & =  (c_2 \id - 6 r k_x \sz) a_y^2(t)  
      + (2 c_2 k_y \id -12 r k_{x} k_y \sz + v \sx) a_y(t)~, \\
    W_{xy} (t)  & =   -6 r \sz a_x(t)a_y^2(t)  -12 r k_y \sz a_x(t) a_y(t)~.
    \label{eq:FQTrig:Vxy}
\end{align}
\end{subequations}
Here the elementary charge $e$ and the reduced Planck constant $\hbar$ have been included in units of the components of the vector potential $a_x$ and $a_y$  to shorten the notation.

We consider a Gaussian pulse given by \eref{eq:LinDir:Gaussian} and set $\eta(t)= e^{-(t/\tau)^2}$. For a fixed $\eta$,  the external driving couples replicas up to third order due to the higher-order contributions of momenta. The Floquet-Fourier Hamiltonian is then given by \eref{eq:FQP:Heff} that can be written as
\begin{equation}
    H^F_{mn} = \left( H_0 - m \hbar \omega \right) \delta_{m,n}
    + Q^{(m-n)}~,
\end{equation}
where the couplings between replicas are given by
\begin{subequations} \label{eq:Bi2Se3:Qs}
\begin{align}
Q^{(0)}  = &
\eta^2 \frac{c_2}{2} (A_x^2 + A_y^2) \id
 + \eta^2 3 r k_x (A_x^2 + A_y^2) \sz ~, \label{eq:Bi2Se3:Q0}\\
    Q^{(1)}  = &
+ \eta c_2 (i A_x k_x + A_y k_y) \id
 + \eta \frac{v}{2}\left( A_y \sx -i A_x \sy \right)
\nonumber \\
& + \eta 3 r \left[ i A_x(k_x^2 - k_y^2)  - 2 A_y k_x k_y \right]\sz +\eta^3 \frac{3ir}{4} A_x (A_x^ 2-A_y^ 2) \sz ~,
\\
Q^{(2)} = & - \eta^2 \frac{c_2}{4} (A_x^2-A_y^2)\id
 -\eta^2\frac{3r}{2} \left[ k_x(A_x^2+A_y^2) + 2 i k_y A_x A_y\right] \sz ~,
\\
Q^{(3)} = & - \eta^3  \frac{i r A_{x}}{4}\left(A_{x}^{2} + 3 A_{y}^{2}\right) \sz ~,\\
    Q^{(-i)} = & \left(Q^{(i)}\right)^{\dagger}~ \text{for }i=1,2,3.
\end{align}
\end{subequations}
Note that, in addition to the coupling between replicas, the pulse also modifies the energy of the bands themselves through $Q^{(0)}$. This term results in a trivial energy shift, proportional to $c_2$, and a coupling from the trigonal warping contribution which tends to close the gap between the bands, proportional to $r$.

The trigonal warping term then generates different phenomena compared to the case of the simple linear Dirac Hamiltonian studied in the previous section~\ref{sub:LinDirac}.
To be specific, the pulse frequency is fixed in the mid-infrared range to $\SI{160}{\milli \eV}$, which corresponds to $\SI{38.7}{\tera \Hz}$. This value 
is taken as a reference from the experiment in \cite{Mahmood2016}. We 
further choose linear polarisation
because it agrees with measurements by minimising the so-called laser-assisted photo emission (LAPE)~\cite{Mahmood2016, Ito2023}. In ARPES experiments, LAPE is due to the dressing of the free electron states near the surface of a solid in a pump-probe setup~\cite{Freericks2009,Sentef2013, Schueler2021a,Schueler2022}. This effect is usually modelled by Bloch states transitioning to Volkov states, which are the solutions of the TDSE for a free electron interacting with an electromagnetic field (see \cite{Park2014} for the derivation of Floquet and Volkov states for Dirac Hamiltonians). Both Floquet and Volkov states exhibit sidebands, and the separation of the two contributions is of primary importance for the correct interpretation of the ARPES intensities. When the driving field is polarised in the surface plane~\cite{Mahmood2016, Ito2023} the Volkov states are minimised. At the same time, as we will show below for the  Bi$_2$Se$_3$ Hamiltonian, the linear polarisation still exhibits sideband transitions due to the higher-order terms in the momenta.

\subsubsection{Linearly polarized Gaussian pulse}

\begin{figure}
        \centering
        \includegraphics[width=0.8\columnwidth]{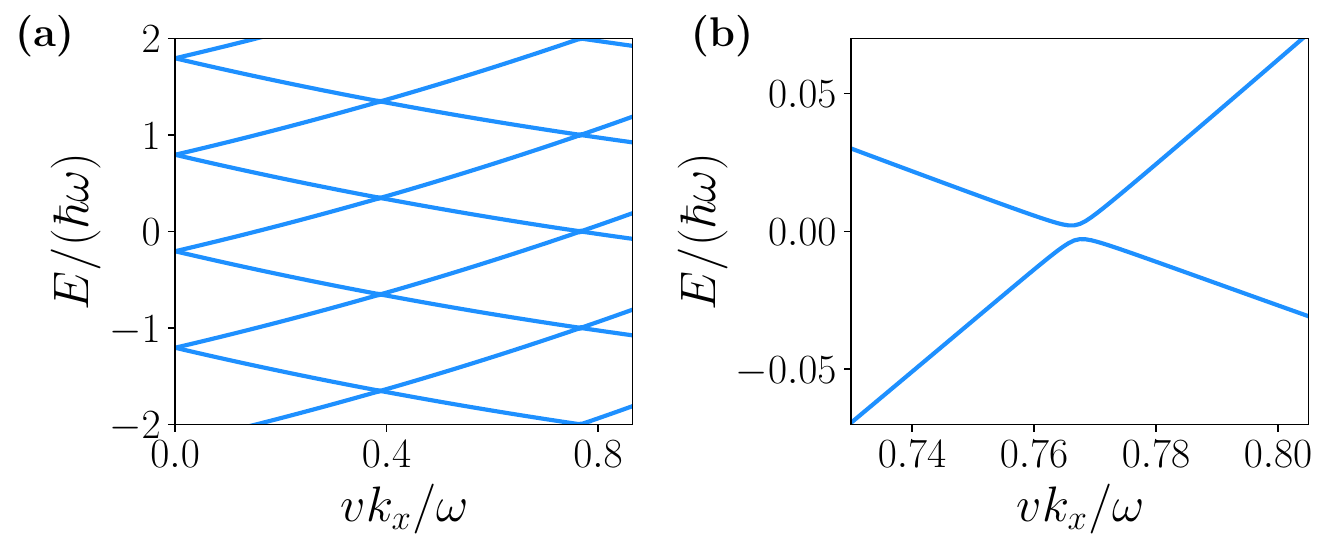}
        \caption{Floquet quasi-energy spectrum for the linearly polarised drive with amplitude $ v e A_x \eta /(\hbar \omega) = 4 $. In panel (a) the dispersion is plotted for a wider range of $k_x$, while in panel (b) the (avoided) crossing near zero energy is zoomed in.}
        \label{fig:Bi2Se3:gap}
    \end{figure}

The linear polarisation in this modified Dirac equation leads to much richer physics than in the previously studied linear Dirac model. In fact, the high-order coupling terms together with the interband coupling $Q^{(0)}$ allow for transitions between replicas from different bands and create band gaps in the linearly polarised case. In contrast to the Dirac Hamiltonian, the commutator
\begin{equation} \label{eq:Bi2Se3:conmut}
    [H_0 (k_x, k_y = 0), W_x(t)] =  -4i v r k_x a_x(t)  (a_x^2(t) + 3a_x(t)k_x + 2k_x^2)~\sx~,
\end{equation} 
is non-zero and contains higher-order terms in the momentum and vector potential leading to  couplings between the replicas.
For the drivings considered here, the (avoided) crossings of the replicas have gaps of less than $\SI{10}{\milli \eV}$.
Figure~\ref{fig:Bi2Se3:gap} shows
a zoom into a gapped Floquet spectrum for $v e A_x/(\hbar \omega) = 4$. 

\begin{figure}
    \centering
    \includegraphics[width=0.8\columnwidth]{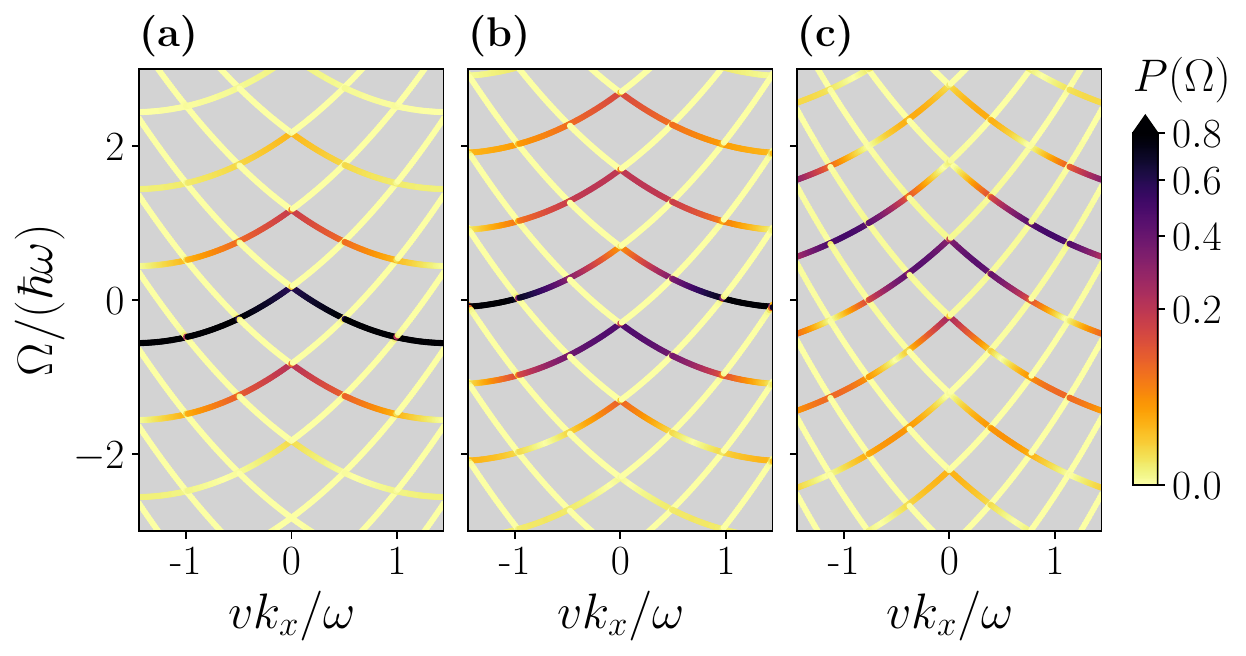}
        \caption{Transition probability amplitude $P(\Omega)$ projected over the Floquet spectrum at $k_y=0$ in the case of linearly polarised driving of a  Hamiltonian representing topological insulator surface states.
        The pulse amplitudes are $ev A_x \eta /(\hbar \omega) = \{1, 2, 4 \}$ in panels (a), (b) and (c), respectively. The colour code corresponds to the amplitude of $P(\omega)$, calculated according to \eref{eq:FQP:P}. The initial state $\psi_0$ used in the calculations is the valence state for the Hamiltonian~\eref{eq:Bi2Se3:H0} at $k_y=0$. }
        \label{fig:Bi2Se3:Pomega}
\end{figure}

In \fref{fig:Bi2Se3:Pomega}, the transition probability amplitude $P(\Omega)$ computed from \eref{eq:FQP:P} is projected over the Floquet spectrum for fixed $\eta$. In contrast to the linear Dirac model shown in \fref{fig:LinDirac:P}, $P(\Omega)$depends on $k_x$ for Bi$_2$Se$_3$. The initial state chosen in \fref{fig:Bi2Se3:Pomega} is once more the valence band state of $H_0$ leading to a $P(\Omega)$ which spreads mainly over valence band replicas. The number of replicas involved depends on the pulse strength: in panel~(a) the original band is mainly occupied, with a smaller contribution in the first and second replicas, while for $ev A_x \eta /(\hbar \omega) = 4$ in panel~(c) $P(\Omega)$ spreads over sidebands of different orders, depending on $k_x$.

\begin{figure}
        \centering
        \includegraphics[width=0.8\columnwidth]{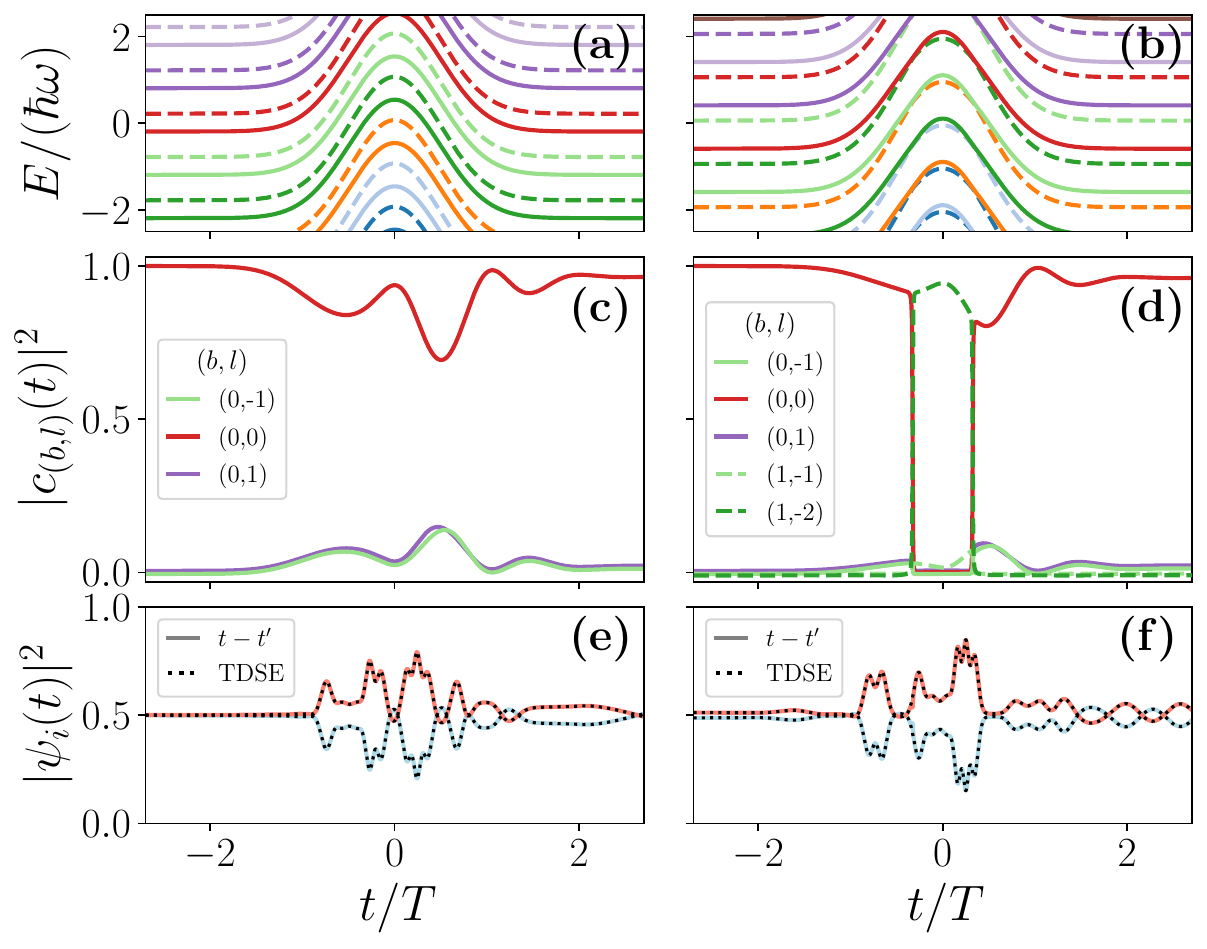}
        \caption{Floquet quasi-energies (a,b), expansion coefficients (c,d) and evolution of wavefunction components (e,f) as a function of time for $vk_x/\omega =0.21 ~(0.82)$ in the left (right) panels.
        The other parameters are fixed at $k_y=0$, $\tau/T = 1$ and $e v A_i/(\hbar\omega) = 4$.
        In (a,b) the colour code distinguishes the quasi-energies with from the same replica and the valence (conduction) band is denoted by a continuous (dotted) line.
        To simplify the plot, in (c,d) only the non-zero expansion coefficients are plotted, corresponding to the replicas indicated in the legend. The notation  employed is $(b,l)$, where $b$ is the band index and $l$ is the replica index.
        In (e,f) a perfect match is found by comparing the direct solution of the TDSE and the solution based on the $t-t^{\prime}$ formalism.
        }
        \label{fig:Bi2Se3:calpha}
\end{figure}

We fix the momenta to two representative values $vk_x/\omega =0.21 ~(0.82)$, corresponding to the left (right) columns of \fref{fig:Bi2Se3:calpha}. The state at $vk_x/\omega =0.21$ is near to the Dirac point and hence far from any replica crossing when the pulse is included. On the other hand, the $vk_x/\omega =0.82$ state is right next to the hybridized gap between replicas of second order, see \fref{fig:Bi2Se3:gap}.
The initial state considered is the valence eigenstate of the unperturbed Hamiltonian, corresponding to $(b,l)= (0,0)$, 
so that $c_{(0,0)} = 1$ at $t \to - \infty$. Due to the trigonal warping and the quadratic terms, the Floquet quasi-energies are modified by the external pulse. For $vk_x/\omega =0.21$ in panel~(a) the Floquet quasi-energies follow the rising and decaying of the pulse envelope, while for $vk_x/\omega =0.82$ in panel~(d) the conduction and valence bands hybridise at two different times.

\par For the state closer to the Dirac point, in the left panels of \fref{fig:Bi2Se3:calpha}, the evolution is dictated by a shift of the occupation towards the first valence sidebands, i.e. to $(b,l) = (0,\pm 1)$. On the other hand, for the state at $vk_x/\omega = 0.82$, in the right panels of the \fref{fig:Bi2Se3:calpha} the occupation is shifted towards the conduction sidebands due to the hybridisation of the bands. Note that due to the higher momenta, the hybridising sidebands come from different replicas, in particular the hybridisation occurs mainly between replicas two orders lower, i.e. between $(0,n)$ and $(1, n-2)$.

\begin{figure}
        \centering
        \includegraphics[width=0.8\columnwidth]{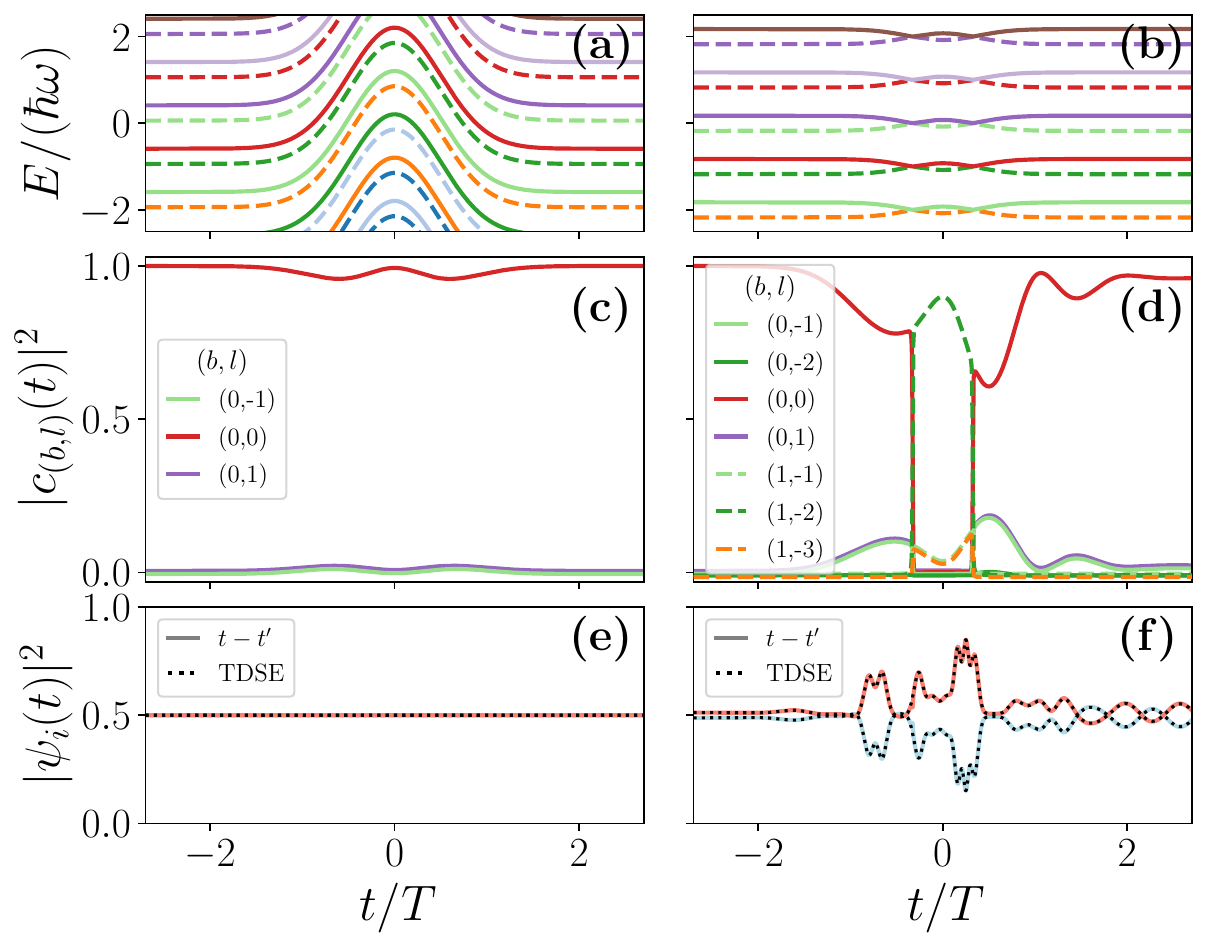}
        \caption{Floquet quasi-energies (a,b), expansion coefficients (c,d) and evolution of wavefunction components (e,f) as a function of time for the quadratic Dirac model and the trigonal warping Dirac model in the left and right panels, respectively.
        The other parameters are fixed to $vk_x/\omega =0.82$, $k_y=0$, $\tau/T = 1$ and $e v A_i/(\hbar \omega) = 4$.
        In (a,b) the colour code distinguishes the quasi-energies with from the same replica and the valence (conduction) band is denoted by a continuous (dotted) line.
        In (c,d) only the non-zero expansion coefficients are plotted, corresponding to the replicas indicated in the legend. The notation  employed is $(b,l)$, where $b$ is the band index and $l$ is the replica index.}
        \label{fig:Bi2Se3:compModels}
\end{figure}

The richer dynamics of the higher $k_x$ modes of the Bi$_2$Se$_3$ states driven by the linearly polarised pulse is  due to the higher-order terms in the Hamiltonian. It is similar to the case of the circularly polarised pulses in the linear Dirac cone. It is therefore interesting to analyse the effect of these terms in more detail. In \fref{fig:Bi2Se3:compModels} the evolution of the mode with $v k_x/\omega =0.82$ is plotted considering two limiting cases: the Dirac model with quadratic onsite corrections, obtained by setting $r=0$ in the Hamiltonian~\eref{eq:Bi2Se3:Ht}, and the Dirac model with trigonal warping corrections, obtained by setting $c_2=0$ in~\eref{eq:Bi2Se3:Ht}. The results are plotted in \fref{fig:Bi2Se3:compModels} for the Dirac model with quadratic corrections and the Dirac model with trigonal warping in the left and right panels, respectively. The quadratic correction produces the most important part of the shift of the quasi-energies with the pulse, due to the contribution of the first term in $Q^{(0)}$ in \eref{eq:Bi2Se3:Q0}, which is proportional to $c_2$. This is clearly visible in panel~\ref{fig:Bi2Se3:compModels}(a). However, the evolution of the Floquet coefficients is almost trivial, with a slight shift of the occupation towards the sidebands $(0, \pm 1)$ [see panel~\ref{fig:Bi2Se3:compModels}(c)]. The obtained evolution is mainly given by a phase which is not visible in the absolute value of the wavefunction plotted in panel~\ref{fig:Bi2Se3:compModels}(e). On the other hand, as shown in the right panels of \fref{fig:Bi2Se3:compModels}, the model with only trigonal warping terms clearly encodes the main part of the sideband evolution. The coupling of the bands is indeed produced by this term, which affects both the quasi-energies [see panel~\ref{fig:Bi2Se3:compModels}(b)], and the expansion coefficients [see panel~\ref{fig:Bi2Se3:compModels}(d)].

\begin{figure}
    \centering
    \includegraphics[width=0.8\columnwidth]{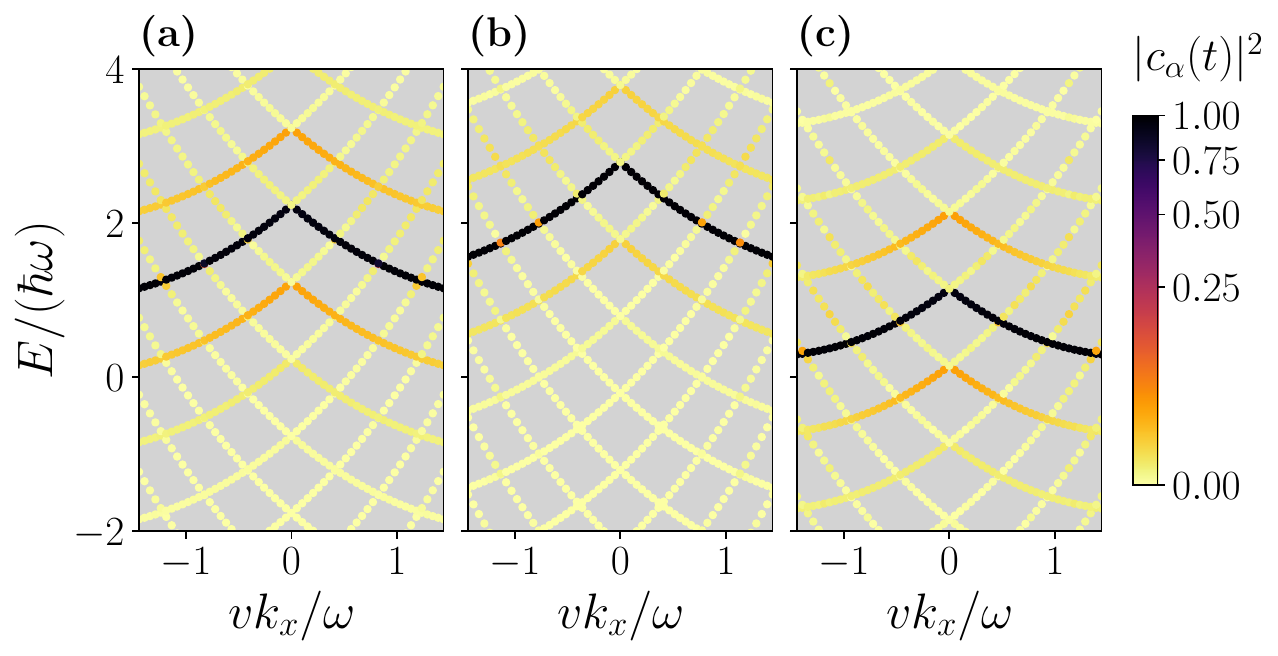}
    \caption{Expansion coefficients $c_\alpha(t)$ projected over the Floquet spectrum for three times $t/T = \{- 0.5, 0, 1 \}$, in panels (a), (b) and (c), respectively.
    The considered pulse is linearly polarised with $\tau/T = 1.5$ and maximum amplitude $ev A_x/(\hbar \omega) = 4$.
    The initial state considered is the valence eigenstate of $H_0(k_x, k_y=0)$.
    The spectrum is plotted for $k_y=0$.}
    \label{fig:Bi2Se3:Spectrum}
\end{figure}

Having elucidated the importance of the trigonal warping term in enabling the transitions between sidebands, we study the full Floquet-Bloch spectrum under linearly polarised Gaussian pulses. The results are plotted in \fref{fig:Bi2Se3:Spectrum}, which shows the projection of the expansion coefficients of the $t-t^{\prime}$ basis over the quasi-energies for three instants of the time evolution corresponding to $t/T = \{- 0.5, 0, 1\}$. Further time snapshots are plotted in \fref{fig:Bi2Se3:Spectrum_6panels}. In the Floquet-Bloch spectrum, the main effect of the driving is indeed the shift of the Dirac cone towards higher energies. In addition, the replicas $(0,\pm 1)$ are populated by the pulse, as expected from \fref{fig:Bi2Se3:calpha}(c), leading to a non-zero amplitude when the pulse is over. Within the full band picture it is also easier to interpret the avoided band crossings generated for $vk_x/\omega = 0.82$ and shown in \fref{fig:Bi2Se3:calpha}(b). These couplings cause indeed the permutation of the valence and conduction bands due to the reshaping of the Dirac cone caused by the trigonal warping term. At the crossing points, an extremely small gap is obtained, as already discussed by the commutator relation~\eref{eq:Bi2Se3:conmut}, and thus the band switch is actually achieved by a change in the band type.

Finally, it is interesting to compare the effect of the linearly polarised pulse in the two models studied, i.e. the linear Dirac and the model for Bi$_2$Se$_3$ surface states, both in the regime of linearly polarised irradiation and with the momentum $k_x$ defined in the same direction of the pulse polarization. In the linear Dirac model, the occupation of the Floquet replicas given by $|c_\alpha(t)|^2$ is independent of $k_x$. This is caused by the fact that the dynamics is only a phase gain, proportional to the external pulse amplitude, which is constant for all momenta in the direction of the polarisation of the light. In the linear Dirac model, the band crossings that appear in the Floquet-Bloch spectrum do not show hybridisation due to the absence of coupling between opposite band replicas. The coupling between replicas is achieved in the linear Dirac model by breaking the collinearity of the momentum with the linear pulse polarisation or by considering a circularly polarised light. 

On the other hand, in the modified Dirac Hamiltonian describing the Bi$_2$Se$_3$ surface states, all band crossings are avoided, i.e. gapped, due to the hybridisation generated by the higher order terms in the momentum. The quadratic correction of the onsite energy produces a trivial displacement of the bands, while the trigonal warping term couples the two bands creating avoided band crossings for $k_x$ further away from the Dirac point. The occupation of the Floquet sidebands is then momentum dependent and $|c_\alpha(t)|^2$ indicates a shift between bands in the case of avoided crossings. However, even if the phenomenology is more complex, the Floquet-Bloch spectrum gives a similar result to the linear Dirac model (for example, compare figures~\ref{fig:LinDirac:Lspectrum} and \ref{fig:Bi2Se3:Spectrum}). In fact, for an initial valence state, in both cases the occupancy is mainly distributed between the originally occupied valence band and the valence sidebands.


\section{Conclusion} \label{sec:Conclusion}

In this paper we have implemented the $t-t^\prime$ formalism for Floquet-Bloch Hamiltonians. Within this formalism, both
the notion of time evolution and the concept of (emerging) Floquet sidebands are merged to describe the time-dependent driving and associated dynamics of Bloch states. In particular, this formalism can be implemented for any model with time-periodic coupling, even if it cannot be easily factorised into an envelope function and a periodic part. Thanks to the standard Fourier expansion employed, the computational cost of the implementation can be minimised by analysing the support of the coupling in the Fourier decomposition. In addition, a clever choice of the time evolution parameter $\eta$ could significantly speed up the calculations, as discussed in the \ref{App:Numerics}.

We have applied this previously derived formalism to two examples:
effective Dirac systems with linear dispersion subject to linearly and circularly polarised pulses, and the modified Dirac model, accounting for trigonal warping and describing the Bi$_2$Se$_3$ surface state. Starting from the results of the periodic driving in the usual Floquet formalism, short Gaussian pulses are described within the extended $t-t^\prime$ basis, yielding good agreement with corresponding results from the direct integration of the TDSE. The linear Dirac model is analysed as a valuable simple but fundamental example that can be directly integrated in the case of linear polarisation. The $t-t^{\prime}$ formalism then offers a simple interpretation of the phase acquisition due to the external pulse in terms of transitions between sidebands of the same band type, independent of the momentum. On the other hand, the case of circularly polarised pulses in the linear Dirac model is described by couplings and corresponding transitions between radiation-dressed conduction and valence sidebands.

In the case of Bi$_2$Se$_3$ surface states, already the linearly polarised pulse generates couplings between the valence and conduction bands leading to avoided crossings. In the vicinity of these gaps, the evolution within the expansion coefficient picture leads to a time-dependent shift of the occupancy between replicas from opposite bands. By solving the Floquet-Bloch spectrum within the $t-t^{\prime}$ formalism, it is then possible to interpret the evolution as a redistribution of the occupancy to the upper and lower sidebands.

In conclusion, the results suggest that the $t-t^\prime$ method is a powerful tool for interpreting the dynamics under pulsed ultrafast periodic driving in an auxiliary Floquet basis. In particular, it is possible to compute the Floquet-Bloch spectrum and to interpret the time evolution of the states as a time-dependent occupation of the Floquet sidebands. This allows a more direct comparison with tr-ARPES experiments while taking into account the underlying physics of mapping the Floquet-Bloch spectra as a function of time.
In particular,  we provide a systematic and quantitative analysis of the emergence and dynamical formation of Floquet sidebands for the topological insulator Bi\textsubscript{2}Se\textsubscript{3}, in line with very recent experiments demonstrating the built-up of Floquet-Bloch bands at topological insulator surface states~\cite{Ito2023}.
While in these experiments other light-matter interaction effects may play a role, the $t-t^{\prime}$ Floquet formalism, as described in our work, should be very helpful to unravel the essential physics. 

\ack

We thank Rupert Huber for inspiring conversations.
Work at Madrid has been supported by Comunidad de Madrid (Recovery, Transformation and Resilience Plan), NextGenerationEU from the European Union (Grant MAD2D-CM-UCM5) and “Talento Program” (Grant 2019-T1/IND-14088), Agencia Estatal de Investigación (Grant PID2022-136285NB-C31).
Research at the University of Regensburg has been funded by the
Deutsche Forschungsgemeinschaft (DFG, German Research Foundation) within Project-ID 314695032 – SFB 1277 (project A07).


\appendix

\section{Expansion of the Floquet-Fourier Hamiltonian} \label{App:FQP:Expansion}

As long as the expansion coefficients $f_b$ are determined by the projection over the basis and the driving frequency $\omega$ is a known quantity, the solution of the TDSE~\eref{eq:FQP:TDSE} is reduced to the determination of the Floquet states~$\ket{u_b(t)}$ and the quasi-energies $\xi_b$.

A common strategy for this is to exploit the periodic properties of $\ket{u_b(t)}$ and perform a discrete Fourier  decomposition in terms of the harmonics of the driving frequency as
\begin{equation}
    \label{eq:FQP:u_bm}
        \ket{u_b (t)} = \sum_{m= -\infty}^\infty e^{-im\omega t} ~\ket{u_b^{(m)}}~,
\end{equation}
with $\ket{u_b^{(m)}}$ the $m-$th Fourier coefficient.
Using \eref{eq:FQP:linearcombF} as an \textit{ansatz} of the TDSE~\eref{eq:FQP:TDSE}, the following expression is obtained as a function of the Fourier coefficients
\begin{equation}
    \label{eq:FQP:Heff0}
    \xi_b ~\ket{u_b^{(m)}} = 
     \sum_{n} \left[
    \frac{1}{T} \int_T \dt \ham_\mathrm{per}(t) e^{i(m-n)\omega t}
    - m \hbar \omega \delta_{m,n}
    \right] \ket{u_b^{(n)}}~,
\end{equation}
whith
$\delta_{m,n} = 1/T \int_T dt~ e^{i(m-n)\omega t}$.
The former expression can be interpreted as an eigenvalue equation in the Fourier space as~\cite{Giovannini2019}
\begin{equation} 
    \label{eq:FQP:Heff}
    \xi_b~\ket{u_b^{(m)} } = \sum_{n} H_{mn}^\mathrm{F} ~\ket{u_b^{(n)}}~,
\end{equation}
where $H_{mn}^\mathrm{F}$ corresponds to the effective Floquet Hamiltonian defined by the quantity inside the square brackets in \eref{eq:FQP:Heff0}.
Note that there is no time dependence in the effective Floquet Hamiltonian due to the integral over one period.

By writing the effective Hamiltonian in matrix form, the $H^\mathrm{F}$ is represented by an infinite matrix of $d \times d$ blocks, where $d$ is the size of the Hilbert space of $\ham_\mathrm{per}(t)$.
In this matrix, the diagonal blocks are given by $\ham_0$ with a shift of $- m \hbar \omega$, while the upper and lower diagonal blocks are the Fourier transform terms of $V(t)$.

Although the Fourier series considers infinite modes, the fact that each $\ket{u_b^{(m)}}$ has a support on a limited range of Fourier modes allows the truncation of $H^\mathrm{F}$ to a finite number of Fourier harmonics.
The accuracy of the truncation depends on the Fourier transform of $V(t)$ as well as on the localization of the states in Fourier space.
However, by increasing the size of the Fourier space, it is possible to obtain an accurate result for the first Brillouin zone.




\section{Comments on equation~\eref{eq:FQP:P}} \label{App:P}

The photoelectron spectroscopy is a pump-probe experimental setup, in which an intense radiation \textit{pumps} the system into an excited states that, after a delay time, is subjected to a weak \textit{probe} pulse.
The photo-electrons generated by this second pulse are then detected with energy and angle resolution.

The photoelectron spectroscopy intensity is related to the transition probability between
a scattering photo-electron state $\ket{\chi_{{\bm p}}(t_f)}$ with momentum ${\bm p}$ at $t_f$ and a given state $\ket{\Psi_0(t_i)}$ at time $t_i$
such that the transition matrix element is
\begin{equation}
    \c{M}_{{\bm p}}(t_f,t_i) = \braket{\chi_{{\bm p}}(t_f) |~ U^\mathrm{driving}(t_f,t_i)~}{\Psi_0(t_i)}~,
\end{equation}
where $U^\mathrm{driving}$ describes the evolution operator of the external fields, in this case the pump and probe drivings~\cite{Giovannini2019}.
The transition probability is then $P({\bm p}) = |\c{M}_{{\bm p}}(t_f,t_i)|^2$ and the intensity $I$ is proportional to such quantity.

\par If the effect of the probe on the states is neglected, the former expression leads to the simplified case of
\begin{equation}
    P(\Omega) = \left| {\int_{-\infty}^{\infty} dt~ e^{i\Omega t} \braket{\psi(t)}{\psi(0)}} \right|^2~,
    \end{equation}
that is the Fourier transform of the projection on the initial state $\ket{\psi(0)}$ of the solution of the TDSE $\ket{\psi(t)}$, considering only the pump driving.


\section{Derivation of equation~\eref{eq:FQk:Gab}} \label{App:eqGab}
For simplicity we omit the dependency on the momentum ${\bm k}$ in the following expressions.
On the other hand, the dependence of $\eta(t)$ is going to be explicitly indicated: if $\eta(t)$ means the time-dependence is considered, otherwise $\eta$ is going to be treated as a parameter.
The starting point is the TDSE formulated for the Hamiltonian $\widehat{H}(\eta(t),t)$ in \eref{eq:App:TDSE}, the decomposition given by \eref{eq:FQk:psit_dec} and the general property of the Floquet states, equation \eref{eq:App:phiF}:

\begin{align}
    i \hbar \frac{d\ket{\psi(t)}}{dt}~ = \widehat{H}(\eta(t), t)~\ket{\psi(t)}~,\label{eq:App:TDSE}\\
    i \partial_t \ket{u_\alpha(\eta, t)} = \left[ -\xi_\alpha(\eta) + \widehat{H}(\eta, t)\right] \ket{u_\alpha(\eta, t)} \label{eq:App:phiF}
\end{align}
Combining those two equations it is obtained:
\begin{equation} \label{eq:App:c_a_inter}
    i \hbar \frac{d c_{\alpha}}{dt} =
    \xi_\alpha(\eta(t)) c_{\alpha}
    - i \frac{\partial \ket{u_\alpha (\eta(t), t)}}{\partial \eta} \frac{d \eta (t)}{d t} c_\alpha ~,
\end{equation}
Next, the last term of \eref{eq:App:c_a_inter} is arranged in a more compact form employing the completeness relation of the Floquet states following the same steps of~\cite{Ikeda2022}
\begin{subequations}
\begin{align}
    - i  \frac{\partial \ket{u_\alpha (\eta(t), t)}}{\partial \eta} &  = \sum_\beta \int_0^T \frac{dt'}{T}
    \ket{u_\beta (\eta(t), t)} \bra{u_\alpha (\eta(t), t')} \partial_\eta \ket{u_\alpha (\eta(t), t')}~, \\
    & = \sum_{\beta} \widehat{G}^{tt^\prime}_{\alpha \beta} ({\bm k},\eta(t)) \ket{u_\beta (\eta(t), t)}~,
\end{align}    
\end{subequations}
where we have defined
\begin{equation} \label{eq:App:hatGv1}
    \widehat{G}^{tt^\prime}_{\alpha \beta} ({\bm k},\eta) \equiv
    \int_{0}^T \frac{dt'}{T} \braket{u_\alpha({\bm k}, \eta,t')}{\partial_\eta u_\beta({\bm k}, \eta, t')}~.
\end{equation}
\par Finally, employing the definition of the Fourier modes given by \eref{eq:FQk:u_bm} we can re-write \eref{eq:App:hatGv1} as:
\begin{equation}
        \widehat{G}^{tt^\prime}_{(b,l)(b',l')}({\bm k},\eta)
    = \sum_{m} \braket{u_{b}^{(m+l-l')}({\bm k}, \eta)}{\partial_\eta u_{b'}^{(m)}({\bm k}, \eta)}~.
\end{equation}
where we have used again $\delta_{m,n} = 1/T \int_T dt~ e^{i(m-n)\omega t}$.


\section{Numerical implementation} \label{App:Numerics}
This Appendix is devoted to brief comments on the numerical implementation of the $t-t'$ formalism. The full code is available in the repository~\cite{github}. 

\par The first important point, already emphasised in the text, is the availability of an (analytical) insight into the Fourier expansion.
This allows a minimal expansion of the Hilbert space in the Fourier modes for the numerical implementation.
Compared to the direct solution of the TDSE, the $t-t'$ formalism indeed enlarges the Hilbert space required by the Fourier expansion. However, the size of the Floquet-Fourier space can be easily controlled by computing the Fourier expansion of the pulse-induced couplings in the Hamiltonian, the $W_i$ terms in \eref{eq:LinDir:H_W} and~\eref{eq:Bi2Se3:Ws}. In this way, the support on the Fourier replicas of the eigenvectors of the Floquet-Fourier Hamiltonian can be easily identified by the number of replicas coupled by the $Q_i$ terms, Eqs.~\eref{eq:LinDir:Qop} and~\eref{eq:Bi2Se3:Qs}.

\par Apart from the size of the Hilbert space of the Floquet-Fourier expansion, since the evolution is written in terms of the parameter $\eta$, it is indeed possible to optimise the numerical calculations by considering $\eta(t)$ functions that lead to the same $\eta$ values for different times and maximum amplitudes. In fact, one of the more expensive parts of the calculation is solving the eigenvector problem for the Floquet Hamiltonian, which is done at fixed $\eta$. Nevertheless, different $\eta(t)$ functions can lead to the same values of $\eta$, so the diagonalisation can be reused for a different set of parameters if $\eta(t)$ is conveniently defined.

\par The numerical implementation employed in this work favours the analytic insight and the pedagogical approach more than the numerical cost-effectiveness. The code is written in \texttt{python}, mainly employing \texttt{NumPy} built-in functions  and the symbolic calculations of the Fourier expansions are done in \texttt{SymPy}~\cite{sympy2017}. However, the code provided can be easily tailored for the numerical optimization of a specific Hamiltonian. In particular, the symbolic part can be speeded up more directly by writing the \texttt{NumPy} functions needed for the specific model under consideration without using symbolic integral expressions.
The code is structured in classes that match the main definitions of the article, the correspondence between the Hamiltonians defined in the text and the code classes are the following:
\begin{itemize}
    \item \texttt{Hamiltonian} takes as input a symbolic Hamiltonian and the terms of the vector potential in symbolic form. It corresponds to the definition of the Hamiltonian $\ham_\mathrm{per}(t)$ in \eref{eq:FQP:Ht}.
    Its methods \texttt{fourier\_elements}(n) and \texttt{fourier\_elements\_lambify} compute, for a fixed amplitude and parameter set, the Fourier expansion elements $Q_i$ for $i=1,..n$ in a symbolic expression and a \texttt{NumPy} function form, respectively.
    The time evolution operator $U(t)$ of \eref{eq:U_t} is defined by the method \texttt{time\_evolutionU}.
    \item \texttt{Hamiltonian\_FloquetFourier} takes as input a \texttt{Hamiltonian} class and corresponds to the evaluation of the Fourier expansion of the Floquet Hamiltonian.
    The method \texttt{fourier\_hamiltonian} returns the Floquet-Fourier Hamiltonian $H_{mn}$ from \eref{eq:FQP:H_FF} in matrix form and \texttt{fourier\_spectrum} evaluates the Floquet spectrum given by \eref{eq:FQttp:xi_bl}.
    \item \texttt{Hamiltonian\_ttp} is the class for defining the $t-t'$ Hamiltonian according to \eref{eq:FQk:Http}.
    The method \texttt{ifs\_basis} computes the instantaneous Floquet basis for the $t-t'$ decomposition in \eref{eq:FQttp:psit_dec} and \texttt{ifs\_Chamilt} corresponds to the Hamiltonian of the coefficients $c_\alpha(t)$ in \eref{eq:FQttp:Cham}.
\end{itemize}

Apart from the classes for the Hamiltonian expressions, there are two main auxiliary classes:
\begin{itemize}
    \item \texttt{ObservablesFloquet} evaluates some relevant observables in the static Floquet picture, such as the time-averaged density of states with \texttt{timeAveragedDOS} and the photoelectron spectroscopy intensity (\texttt{photoelAmp}) according to \eref{eq:FQP:P}.
    \item \texttt{IFS\_solver} solves the evolution of the $c_\alpha(t)$ coefficients of the $t-t'$ formalism by integrating the differential equation with the method \texttt{c\_t} and labelling the replicas according to the convention in \eref{eq:FQk:limEk} in \texttt{tag\_fqlevels}.
\end{itemize}
%


\section{Supplemental figures} \label{App:ExtraFigs}

Here we provide additional figures referred to in the main text, representing further parameter regimes.
\begin{figure}[htb]
        \centering
        \includegraphics[width=0.8\columnwidth]{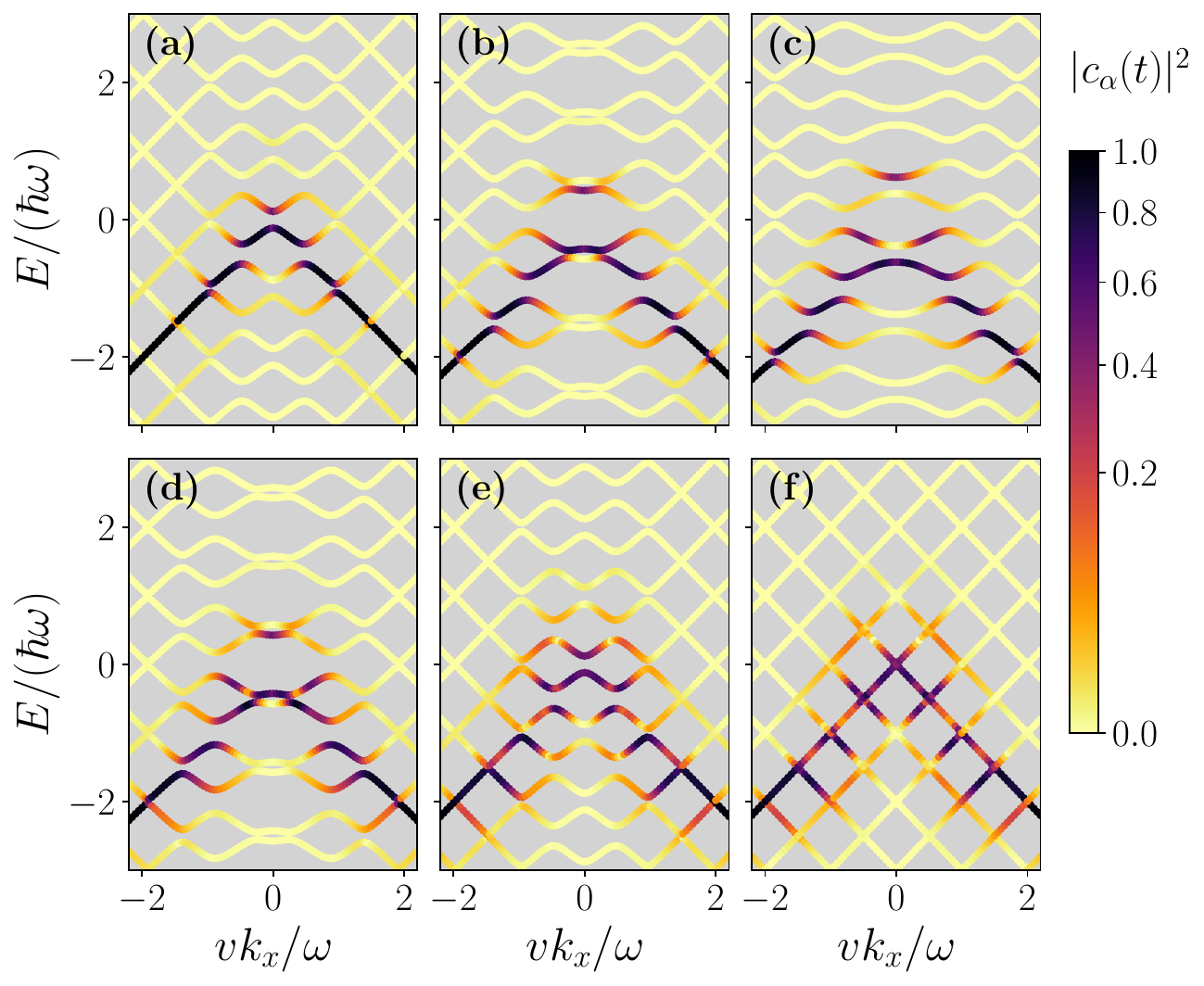}
        \caption{Expansion coefficients $c_\alpha(t)$ projected over the Floquet spectrum for three instants of time $t/T = \{-1, - 0.5, 0, 0.5, 1, 2\}$, in panels (a) to (f).
        The pulse considered is circularly polarized pulse with $\tau/T = 1$ and maximum amplitude $ev A_x/(\hbar \omega) = 1$.
        The initial state considered is the valence eigenstate of $H_0(k_x, k_y=0)$.
        The spectrum is plotted for $k_y=0$.}
        \label{fig:LinDirac:Cspectrum_6panels}
        \end{figure}
        
\begin{figure}[htb]
    \centering
    \includegraphics[width=0.8\columnwidth]{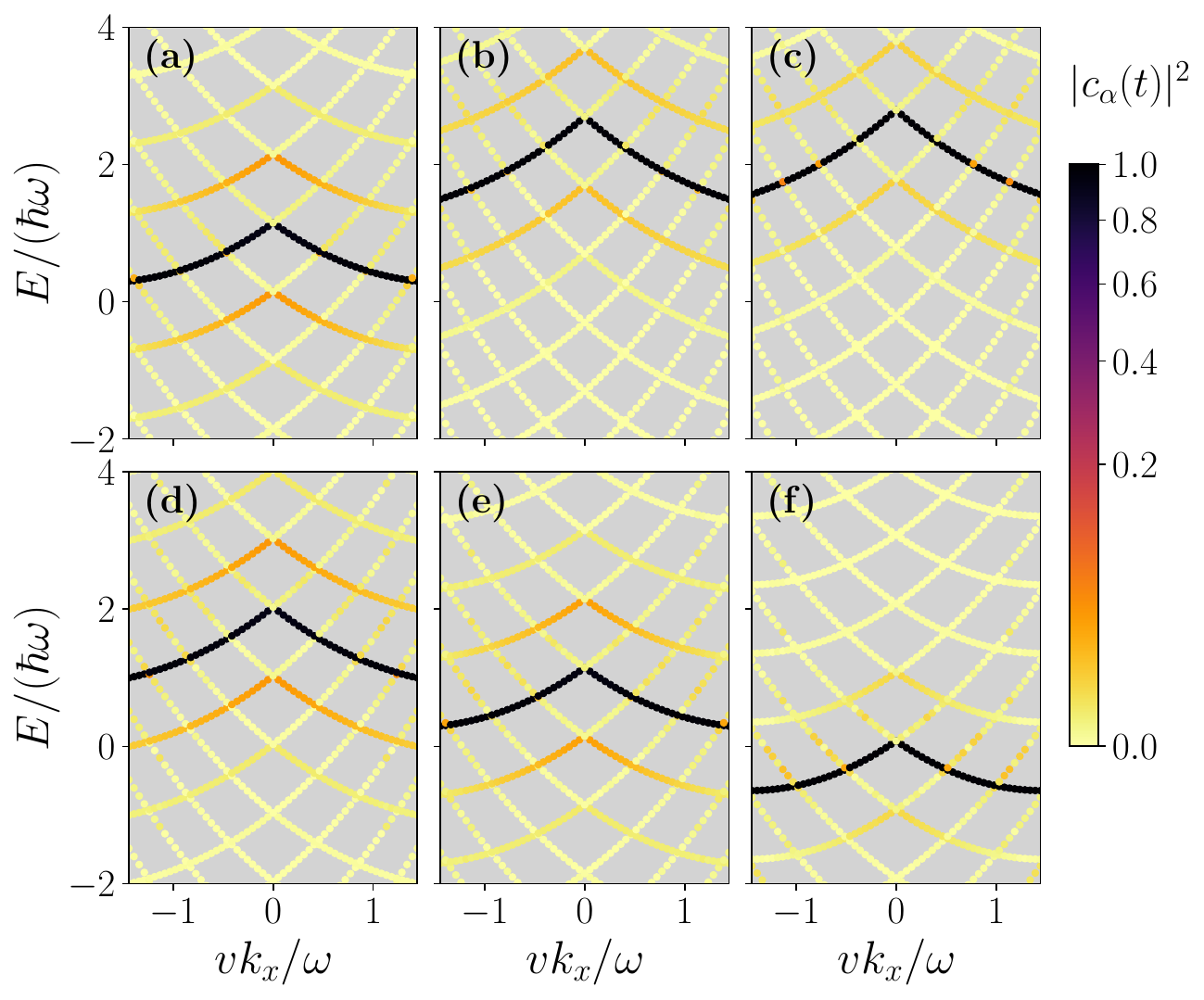}
    \caption{Expansion coefficients $c_\alpha(t)$ projected over the Floquet spectrum for three instants of time $t/T = \{-1, - 0.2, 0, 0.6, 1, 2\}$, in panels (a) to (f).
    The pulse considered is linearly polarized with $\tau/T = 1.5$ and maximum amplitude $ev A_x/(\hbar \omega) = 4$.
    The initial state considered is the valence eigenstate of $H_0(k_x, k_y=0)$.
    The spectrum is plotted for $k_y=0$.}
    \label{fig:Bi2Se3:Spectrum_6panels}
\end{figure}


\end{document}